\documentclass[11pt,a4paper]{article}

\usepackage{jheppub}
\usepackage{graphicx}
\usepackage{xspace}
\usepackage{booktabs}

\usepackage{makecell}

\usepackage{xspace}

\usepackage{adjustbox}

\usepackage{bm}

\usepackage{multirow}

\makeatletter
\g@addto@macro\bfseries{\boldmath}
\makeatother

\newcommand{\ttbar}{t\bar{t}}

\newcommand{\rf}{\text{ref}}

\newcommand{\GeV}{\,\mathrm{GeV}}
\newcommand{\TeV}{\,\mathrm{TeV}}

\newcommand{\as}{\alpha_s}

\newcommand{\yes}{\checkmark}  
\newcommand{\no}{$-$}

\newcommand{\arxiv}[1]{\href{http://arxiv.org/abs/#1}{arXiv:#1}}
\newcommand{\arXiv}[1]{\arxiv{#1}}

\definecolor{darkgreen}{rgb}{0,0.5,0}
\definecolor{darkblue}{rgb}{0,0,0.7}
\definecolor{darkred}{rgb}{0.5,0,0.0}
\definecolor{darkorange}{rgb}{0.8,0.4,0.0}

\newcommand{\stt}{\sigma_{t\bar{t}}}
\newcommand{\asmz}{\alpha_s \left( m_Z \right) }
\newcommand{\mt}{m_t}

\newcommand{\cell}[2]{ \makecell{\parbox{#1}{\centering\strut #2}} }
\newcommand{\lcell}[2]{ \makecell{\parbox{#1}{\strut #2}} }

\newcommand{\alphasResultCenter}{0.1177}        
\newcommand{\alphasResultLeftError}{0.0036}     
\newcommand{\alphasResultRightError}{0.0034}    
\newcommand{\alphasResultSymmPercentage}{3.0}   

\title{Determination of the strong coupling constant $\asmz$ from measurements of the
  total cross section for top-antitop quark production}

\author[a]{Thomas Klijnsma,}
\author[b]{Siegfried Bethke,}
\author[a]{G\"unther Dissertori,}
\author[c,*]{Gavin P. Salam\note[*]{On leave from CNRS, UMR 7589, LPTHE, F-75005, Paris, France},}
\affiliation[a]{Institute for Particle Physics, ETH Zurich, Zurich, Switzerland}
\affiliation[b]{Max-Planck-Institute of Physics, Munich, Germany}
\affiliation[c]{CERN, Theoretical Physics Department, CH-1211 Geneva 23, Switzerland}

\preprint{CERN-TH-2017-183}


\abstract{
    We present a determination of the strong coupling constant $\asmz$ using
    inclusive top-quark pair production cross section measurements performed at the
    LHC and at the Tevatron.
    Following a procedure first applied by the CMS collaboration,
    we extract individual values of $\asmz$ from measurements by different
    experiments at several centre-of-mass energies, using
    QCD predictions complete in NNLO perturbation theory, supplemented
    with NNLL approximations to all orders, and suitable sets of parton
    distribution functions.
    The determinations are then combined using a likelihood-based
    approach, where special emphasis is put on a consistent treatment
    of theoretical uncertainties and of correlations between various
    sources of systematic uncertainties.
    Our final combined result is
    $\asmz =
    \alphasResultCenter^{+\alphasResultRightError{}}_{-\alphasResultLeftError{}}$.
}

\begin{document}
\maketitle

\section{Introduction}

The strong coupling constant of Quantum Chromodynamics (QCD), $\as$,
is, together with the quark masses, the main free parameter of the QCD
Lagrangian. 
It enters into every process that involves the strong interaction and
is the fundamental parameter of the perturbative expansion used in
calculating cross sections for processes with large momentum
transfers.

The strong coupling is a function of a renormalisation scale $\mu$.
Its dependence on $\mu$ is governed by renormalisation group
equations~\cite{Baikov:2016tgj,Herzog:2017ohr}, however its value at a given reference
scale must be determined from experimental data.
The current world average value for the coupling evaluated at the
$Z$-boson mass scale, $\asmz$, as determined by the Particle Data
Group (PDG), is $0.1181 \pm 0.0011$~\cite{PDG}.
The world average incorporates information from a wide variety of
experimental data and of methods to deduce $\as$ from that data.
It requires at least next-to-next-to-leading order (NNLO) accuracy in
the perturbative expansions that are used.

Even with the $1\%$ precision that is quoted by the PDG, the
uncertainty on $\as$ contributes significantly to uncertainties on
physical predictions for colliders.
For example, it leads to about $2\%$ uncertainty on the gluon-fusion
Higgs cross section, comparable with the largest of any of the other
individual uncertainties~\cite{Anastasiou:2016cez}.
Furthermore, while the bulk of the evidence points to values of the
strong coupling that are compatible with $\asmz \simeq 0.118$,
including precise lattice-QCD based
determinations, e.g.~\cite{PDG,Aoki:2016frl,McNeile:2010ji,Bruno:2017gxd},
there are a handful determinations with small quoted uncertainties
that suggest $\asmz$ values that are several standard deviations below
the world average. 
Notable cases are those from the
Thrust and C-parameter distributions in $e^+e^-$ collisions, which
yield $0.1135 \pm 0.0011$ and $0.1123 \pm 0.0015$
respectively~\cite{Abbate:2010xh,Hoang:2015hka},\footnote{ An
  alternative analysis of the Thrust quotes a significantly larger
  uncertainty, $0.1137^{+0.0034}_{-0.0027}$~\cite{Gehrmann:2012sc}.}
or the ABMP PDF fit~\cite{Alekhin:2017kpj}, $0.1147\pm0.0008$.

Of the various NNLO determinations of the strong coupling, so far only one is
based on hadron collider data, using a measurement of the top-quark
pair production cross section ($\stt$) performed by the CMS
Collaboration at a centre-of-mass energy
$\sqrt{s}=7\,$TeV~\cite{CMS-ttbar-alphas}.
It yields $\asmz = 0.1151^{+0.0028}_{-0.0027}$.
This extraction is intriguingly placed between the world average and
the outlying low $\as$ extractions, albeit compatible with both.
However, it is based on a single, early and now outdated measurement of $\stt$.
It is of interest, therefore, to examine how it is affected by more
recent precise measurements by the ATLAS and CMS Collaborations at
CERN's Large Hadron Collider (LHC)
\cite{CMS78TeV,CMS13TeV,CMS13TeVoneleptononejet,ATLAS78TeV,ATLAS13TeV}
as well as by a combination of measurements from the D0 and CDF
collaborations at the Tevatron~\cite{D0CDFcombination}.

In the course of our discussion, we will encounter issues related to
the treatment of theoretical uncertainties and the choice of the
parton distribution function (PDF) set
that are of relevance more generally in the determination of the
strong coupling and other fundamental constants (e.g.\ the top-quark
mass) from collider data.
Such studies may become increasingly widespread in the coming
years, given the recent rapid progress in NNLO calculations, e.g.\ for
vector-boson (e.g.\ Refs.~\cite{Boughezal:2015ded,Ridder:2016nkl}) and
inclusive jet $p_t$ 
distributions~\cite{Currie:2017tfd} at hadron colliders and jet $p_t$
distributions in Deep Inelastic Scattering (DIS)~\cite{Currie:2017tpe}.

\section{Determination of $\as$ from $\ttbar$ cross section measurements}

\subsection{Theory prediction for the top pair production cross section $\stt$}
\label{sec:theory-predictions}

Theory predictions for the dependence of $\stt$ on $\as$ are
calculated using the program
\texttt{top++2.0}~\cite{topplusplus}. 
It provides the computation of the total cross section up to
NNLO~\cite{topplusplus2}, with possible inclusion of soft-gluon
resummation at next-to-next-to-leading logarithmic order (NNLL), as
described in Refs.~\cite{Beneke:2009rj,Czakon:2009zw}.

The predicted cross section is evaluated setting both the
renormalisation scale $\mu_R$ and factorisation scale $\mu_F$ equal to
the top-quark pole mass. 
The theoretical uncertainty associated with missing higher-order
contributions is evaluated by independently varying $\mu_R$ and $\mu_F$
up and down by a factor of 2, under the constraint that
$\frac{1}{2} \leq \mu_R / \mu_F \leq 2$.
The scale uncertainties are modelled as corresponding to a $68\%$
confidence interval with a Gaussian-shaped uncertainty profile.
This choice is more conservative than the (flat) $100\%$ confidence
interval that is sometimes taken for scale variations and used,
notably, in Ref.~\cite{CMS-ttbar-alphas}. 
The latter choice leads to a scale
uncertainty contribution that is smaller by a factor $\sqrt{3}$ (the
ratio of the standard deviations of the two uncertainty profiles).
Note that a $100\%$ confidence
interval for scale uncertainties is known to be inconsistent with the
observation that a significant fraction of NNLO calculations is
outside the scale uncertainty interval of the corresponding NLO
calculation.\footnote{As discussed in \cite{Bagnaschi:2014wea} and
  also \cite{GavinLHCPtalk}. 
  Note that the experience with NLO scale uncertainties may not apply
  to NNLO scale uncertainties.
  In particular, for the two cases of hadron-collider calculations
  available at N3LO accuracy, Higgs production in the
  gluon-fusion~\cite{Anastasiou:2016cez} and
  vector-boson-fusion~\cite{Dreyer:2016oyx} channels, while the
  central NNLO results are outside the NLO scale uncertainty bands, the N3LO
  results are well within the corresponding NNLO bands. }

A further choice that needs to be made is whether to include the NNLL
threshold resummation for the cross section.
This is a procedure that resums terms whose leading-logarithmic (LL)
structure is $(\as \ln^2 N)^n$, where $N \sim d\ln\sigma_{\ttbar}/d\ln s$ and
$s$ is the squared centre-of-mass energy.
When $m_{\ttbar}^2/s$ approaches one, i.e.\ when one approaches the
threshold for $t\bar t$ production, $N$ is proportional to
$1/(1-m_{\ttbar}^2/s$) and the threshold resummation is a necessity.
However, at the LHC and even at the Tevatron, top-pair production is
far from threshold and $N$ is not especially large: for the dominant
gluon-gluon production channel at LHC, $N \simeq 1.4$ for
$m_{t\bar t} = 2 m_t$ and $\sqrt{s} = 7\TeV$; while for the dominant
$q\bar q$ production channel at the Tevatron, $N \simeq 1.8$.
Accordingly, there is debate within the community as to
whether threshold resummation is called for.
On one hand, one may argue that it brings terms that have a certain
physical meaning.
On the other, one may argue that there is no reason why the terms
brought by threshold resummation should dominate over other, neglected
terms, and therefore it is more consistent to include just the
fixed-order contributions, which are known exactly.
We will take an agnostic approach to this question, carry out fits
with and without NNLL resummation, and then average both the central
values and the uncertainties in the two cases in order to obtain our
final result.

The theory prediction for $\stt$ also depends on a choice of PDF set.
Since that choice needs to be related to the data that we fit, we
postpone our discussion of the PDF choice to
section~\ref{sec:pdf-choice}.

\subsection{Measurements of the top pair production cross section}
\label{sec:data}

Our $\as$ determination is performed using seven $\stt$ inputs, listed
in Table~\ref{tab:includedmeasurements}.
The six measurements at the LHC include three updated measurements
by the CMS Collaboration at centre-of-mass energies of 7\,TeV,
8\,TeV~\cite{CMS78TeV} and 13\,TeV~\cite{CMS13TeV}.  
These measurements were performed in the $e \mu$ decay channel,%
\footnote{The $\stt$ measurement by CMS at 13\,TeV using events with
  one lepton and at least one jet in the final
  state~\cite{CMS13TeVoneleptononejet} has a slightly better precision
  than the CMS result used in our analysis. However, the effect on the
  final result is marginal, and using measurements from the same decay
  channel yields a clearer correlation structure for the combination.
}
where the $W$-bosons from the top quark decays each themselves decay
into a charged lepton and a neutrino, one of the $W$ decays producing
an electron, the other producing a muon.
The measurements are based on data collected in the years of 2011, 2012 and 2015
respectively, with integrated luminosities of 5.0\,fb$^{-1}$,
19.7\,fb$^{-1}$, and 2.2\,fb$^{-1}$.
From the ATLAS Collaboration, three similar
measurements performed in the
$e\mu$ decay channel are included, based on datasets with
integrated luminosities of 4.6\,fb$^{-1}$, 20.3\,fb$^{-1}$ and
3.2\,fb$^{-1}$ for the 7\,TeV, 8\,TeV~\cite{ATLAS78TeV} and
13\,TeV~\cite{ATLAS13TeV} centre-of-mass energies
respectively.
A seventh input from
the Tevatron collider~\cite{D0CDFcombination} at a centre-of-mass
energy of 1.96\,TeV is included, which
comprises a combination of measurements performed in multiple decay
channels from both the CDF Collaboration and the D0 Collaboration.  

\newcommand{\xstableWidth}{2cm}
\begin{table*}[ht]
{\footnotesize
\setlength\tabcolsep{0pt}
\renewcommand{\arraystretch}{1.2}
\begin{center}
\begin{tabular}{l c c c c c c }
\toprule
\vbox to 10pt {} &
\cell{1.2cm}{$\stt$ [pb]} &
\cell{\xstableWidth}{Statistical unc. [\%]} &
\cell{\xstableWidth}{Systematic unc. [\%]} &
\cell{\xstableWidth}{Luminosity unc. [\%]} &
\cell{\xstableWidth}{E$_{\text{beam}}$ unc. [\%]} &
\cell{\xstableWidth}{Exp. $\mt$ unc. [\%]} \\
\midrule
ATLAS (7\,TeV)~\cite{ATLAS78TeV}           & $182.5  $ & $1.7 \%$  & $2.3 \%$  & $2.0 \%$  & $0.3 \%$  & ${}^{-0.2\%}_{+0.2\%}$         \\
ATLAS (8\,TeV)~\cite{ATLAS78TeV}           & $242.4  $ & $0.7 \%$  & $2.3 \%$  & $2.1 \%$  & $0.3 \%$  & ${}^{-0.2\%}_{+0.2\%}$         \\
ATLAS (13\,TeV)~\cite{ATLAS13TeV}          & $816.3  $ & $1.0 \%$  & $3.3 \%$  & $2.3 \%$  & $0.2 \%$  & ${}^{-0.3\%}_{+0.3\%}$         \\
CMS   (7\,TeV)~\cite{CMS78TeV}             & $173.4  $ & $1.2 \%$  & $2.5 \%$  & $2.2 \%$  & $0.3 \%$  & ${}^{-0.2\%}_{+0.2\%}$         \\
CMS   (8\,TeV)~\cite{CMS78TeV}             & $244.1  $ & $0.6 \%$  & $2.4 \%$  & $2.6 \%$  & $0.3 \%$  & ${}^{-0.4\%}_{+0.4\%}$         \\
CMS   (13\,TeV)~\cite{CMS13TeV}            & $809.8  $ & $1.1 \%$  & $4.7 \%$  & $2.3 \%$  & $0.2 \%$  & ${}^{-0.8\%}_{+0.8\%}$         \\
TEV   (1.96\,TeV)~\cite{D0CDFcombination}  & $7.52   $ & $2.7 \%$  & $3.9 \%$  & $2.8 \%$  & $0.0 \%$  & ${}^{-1.1\%}_{+1.4\%}$         \\
\bottomrule
\end{tabular}
\end{center}
\caption{\small Cross sections and experimental uncertainties for the
  $\stt$ inputs that we
  use~\cite{D0CDFcombination,CMS78TeV,CMS13TeV,ATLAS78TeV,ATLAS13TeV}.
  The LHC beam energy uncertainties quoted in these references have been scaled down by a factor $6.6$
  in light of the recent beam-energy calibration~\cite{LHCBeamMomentumCalibration}, which has a $0.1\%$
  uncertainty and coincides
  with the nominal energy within uncertainties. 
  The original beam-energy-induced uncertainties corresponded to
  $0.66\%$~\cite{OldBeamEnergy}.
  The Tevatron beam energy uncertainty is sufficiently
  small (cf.\ Ref.~\cite{Johnson:1988bx}) that no beam energy uncertainty is
  quoted by CDF and D0 in the $t\bar t$ cross section measurements.
  The cross section and uncertainties listed here are adjusted to the
  top mass corresponding to the
  latest world average value computed by the Particle Data Group~\cite{PDG},
  $\mt = 173.2 \pm 0.51 \pm 0.71 \GeV$.
  The ``Exp.\ $m_t \text{ unc.}$''
  column corresponds to the $\delta m_t$ 
  uncertainty discussed in section~\ref{sec:top-mass-dependence},
  signed such that the upper (lower) uncertainty corresponds to an
  increase (decrease) in $m_t$.
} 
\label{tab:includedmeasurements}
}
\end{table*}

\subsection{Choice of PDF}
\label{sec:pdf-choice}

Several considerations arise in our choice of PDF. 
Firstly, we restrict our attention to recent global fits that are
available through the LHAPDF interface~\cite{LHAPDF}.
Secondly, we require that the PDFs should be available for at least
three $\as$ values, so that we can correctly determine the $\as$
dependence of the cross section in the context of that PDF.
These two conditions limit us to the CT14~\cite{CT14},
MMHT2014~\cite{MMHT2014} and the NNPDF3.0~\cite{NNPDF30} series.
Thirdly, we impose a requirement that the PDF should not have included
$\stt$ data in its fitting procedure.
As should be obvious qualitatively, and as we will discuss
quantitatively elsewhere~\cite{InPrep}, using a PDF with top-data
included would bias our fits.

\begin{table}
  \centering
  {\setlength\tabcolsep{5pt}
  \begin{tabular}{lccccc}
    \toprule
    & Tevatron        
    & \cell{1.8cm}{ATLAS (7\,TeV)}
    & \cell{1.8cm}{ATLAS (8\,TeV)}
    & \cell{1.8cm}{CMS (7\,TeV)}
    & \cell{1.8cm}{CMS (8\,TeV)}
    \\\midrule
    CT14~\cite{CT14}              & \no  & \no  & \no  & \no  & \no  \\
    MMHT2014~\cite{MMHT2014}      & \yes & \yes & \no  & \yes & \yes \\
    NNPDF30~\cite{NNPDF30}        & \no  & \yes & \yes & \yes & \yes \\
    NNPDF30\_noLHC~\cite{NNPDF30} & \no  & \no  & \no  & \no  & \no  \\
    \bottomrule
  \end{tabular}}
  \caption{Top pair cross section data included in a selection of recent
    PDF fits. A ``\yes'' (``\no'') indicates that the corresponding $t\bar t$
    cross section measurement is (is not)
    included in the PDF fit.
    The specific sets of $7$ and $8\TeV$ ATLAS and CMS data used in
    the fits do not
    always coincide 
    with those that we list in Table~\ref{tab:includedmeasurements}.
    All the PDFs shown here predate the $13\TeV$ measurements.
  }
  \label{tab:pdf-fit-ttbar-choices}
\end{table}

Table~\ref{tab:pdf-fit-ttbar-choices} summarises what data has been
included in each of these PDF sets, including both the default NNPDF30
set and NNPDF30\_nolhc, obtained without LHC data.
One sees that the two options that are available to us are CT14 and
NNPDF30\_nolhc.\footnote{As this article was being completed the
  NNPDF31 series~\cite{Ball:2017nwa} of PDF sets became available. 
  It includes a set fitted without top data, however only
  for a single value of the strong coupling, and accordingly is not
  suitable for use in a strong coupling determination.}

\newcommand{\predxstableWidth}{2.2cm}
\begin{table*}[ht]
{
\footnotesize 
\renewcommand{\arraystretch}{1.3}
\setlength\tabcolsep{3pt}
\begin{center}
\begin{tabular}{l c c c c c }
\toprule
&
\cell{1.8cm}{$\stt^{\text{pred}}(\as^{\text{ref}})$ [pb]} &
\cell{1.8cm}{PDF unc. [\%]} &
\cell{1.8cm}{Scale unc. [\%]} &
\cell{1.8cm}{$\mt$ unc. [\%]} & 
$\displaystyle
    \frac{ \text{d}\ln{\stt(\as^{\text{ref}})} }{ \text{d}\ln\as }
    $ \\
\midrule 
\textbf{CT14 (NNLO)} & & & & & \\
LHC ($7$\,TeV) &              $172.7$  & ${}_{-3.8\%}^{+4.5\%}$ & ${}_{-6.5\%}^{+4.1\%}$ & ${}^{-2.6\%}_{+2.7\%}$ & $2.486$ \\
LHC ($8$\,TeV) &              $246.7$  & ${}_{-3.5\%}^{+4.0\%}$ & ${}_{-6.3\%}^{+3.9\%}$ & ${}^{-2.5\%}_{+2.6\%}$ & $2.404$ \\
LHC ($13$\,TeV) &             $807.3$  & ${}_{-2.7\%}^{+2.6\%}$ & ${}_{-5.6\%}^{+3.5\%}$ & ${}^{-2.3\%}_{+2.4\%}$ & $2.133$ \\
Tevatron ($1.96$\,TeV) &      $7.3$    & ${}_{-2.2\%}^{+3.4\%}$ & ${}_{-5.5\%}^{+3.8\%}$ & ${}^{-2.7\%}_{+2.8\%}$ & $1.757$ \\
\midrule\multicolumn{6}{l}{\textbf{NNPDF30\_nolhc (NNLO)}} \\
LHC ($7$\,TeV) &              $174.8$  & ${}_{-5.0\%}^{+5.0\%}$ & ${}_{-6.5\%}^{+4.1\%}$ & ${}^{-2.6\%}_{+2.7\%}$ & $2.247$ \\
LHC ($8$\,TeV) &              $249.7$  & ${}_{-4.4\%}^{+4.4\%}$ & ${}_{-6.3\%}^{+3.9\%}$ & ${}^{-2.5\%}_{+2.6\%}$ & $2.099$ \\
LHC ($13$\,TeV) &             $816.2$  & ${}_{-2.9\%}^{+2.9\%}$ & ${}_{-5.6\%}^{+3.5\%}$ & ${}^{-2.3\%}_{+2.4\%}$ & $1.681$ \\
Tevatron ($1.96$\,TeV) &      $7.2$    & ${}_{-3.1\%}^{+3.5\%}$ & ${}_{-5.5\%}^{+3.8\%}$ & ${}^{-2.7\%}_{+2.8\%}$ & $2.396$ \\
\midrule\multicolumn{6}{l}{\textbf{CT14 (NNLO+NNLL)}} \\
LHC ($7$\,TeV) &              $177.9$  & ${}_{-3.7\%}^{+4.4\%}$ & ${}_{-3.5\%}^{+2.6\%}$ & ${}^{-2.6\%}_{+2.7\%}$ & $2.545$ \\
LHC ($8$\,TeV) &              $253.6$  & ${}_{-3.4\%}^{+3.9\%}$ & ${}_{-3.5\%}^{+2.6\%}$ & ${}^{-2.5\%}_{+2.6\%}$ & $2.459$ \\
LHC ($13$\,TeV) &             $825.9$  & ${}_{-2.7\%}^{+2.6\%}$ & ${}_{-3.6\%}^{+2.4\%}$ & ${}^{-2.3\%}_{+2.4\%}$ & $2.178$ \\
Tevatron ($1.96$\,TeV) &      $7.4$    & ${}_{-2.2\%}^{+3.5\%}$ & ${}_{-2.9\%}^{+1.6\%}$ & ${}^{-2.7\%}_{+2.8\%}$ & $1.842$ \\
\midrule\multicolumn{6}{l}{\textbf{NNPDF30\_nolhc (NNLO+NNLL)}} \\
LHC ($7$\,TeV) &              $180.1$  & ${}_{-5.0\%}^{+4.9\%}$ & ${}_{-3.5\%}^{+2.6\%}$ & ${}^{-2.6\%}_{+2.7\%}$ & $2.296$ \\
LHC ($8$\,TeV) &              $256.7$  & ${}_{-4.4\%}^{+4.3\%}$ & ${}_{-3.5\%}^{+2.6\%}$ & ${}^{-2.5\%}_{+2.6\%}$ & $2.147$ \\
LHC ($13$\,TeV) &             $835.0$  & ${}_{-2.8\%}^{+2.8\%}$ & ${}_{-3.6\%}^{+2.4\%}$ & ${}^{-2.3\%}_{+2.4\%}$ & $1.722$ \\
Tevatron ($1.96$\,TeV) &      $7.3$    & ${}_{-3.2\%}^{+3.6\%}$ & ${}_{-2.9\%}^{+1.5\%}$ & ${}^{-2.7\%}_{+2.8\%}$ & $2.476$ \\
\bottomrule
\end{tabular}
\end{center}
\caption{\small Predicted cross sections and uncertainties for the
  PDF sets that we use~\cite{CT14,NNPDF30}, as determined with the
  \texttt{Top++} program~\cite{topplusplus} at a reference value
  of $\as^\rf = 0.118$.
  The results are for $m_t = 173.2\GeV$ and the ``$m_t \text{ unc.}$''
  column corresponds to the $\delta m_t$ 
  uncertainty discussed in section~\ref{sec:top-mass-dependence},
  signed such that the upper (lower) uncertainty corresponds to an
  increase (decrease) in $m_t$.
  } 
\label{tab:includedpredictions}
}
\end{table*}

\newcommand{\FitFigureWidth}{0.44}  
\begin{figure}[t]
\centering
\begin{tabular}{ccc}
\includegraphics[width=\FitFigureWidth\linewidth]{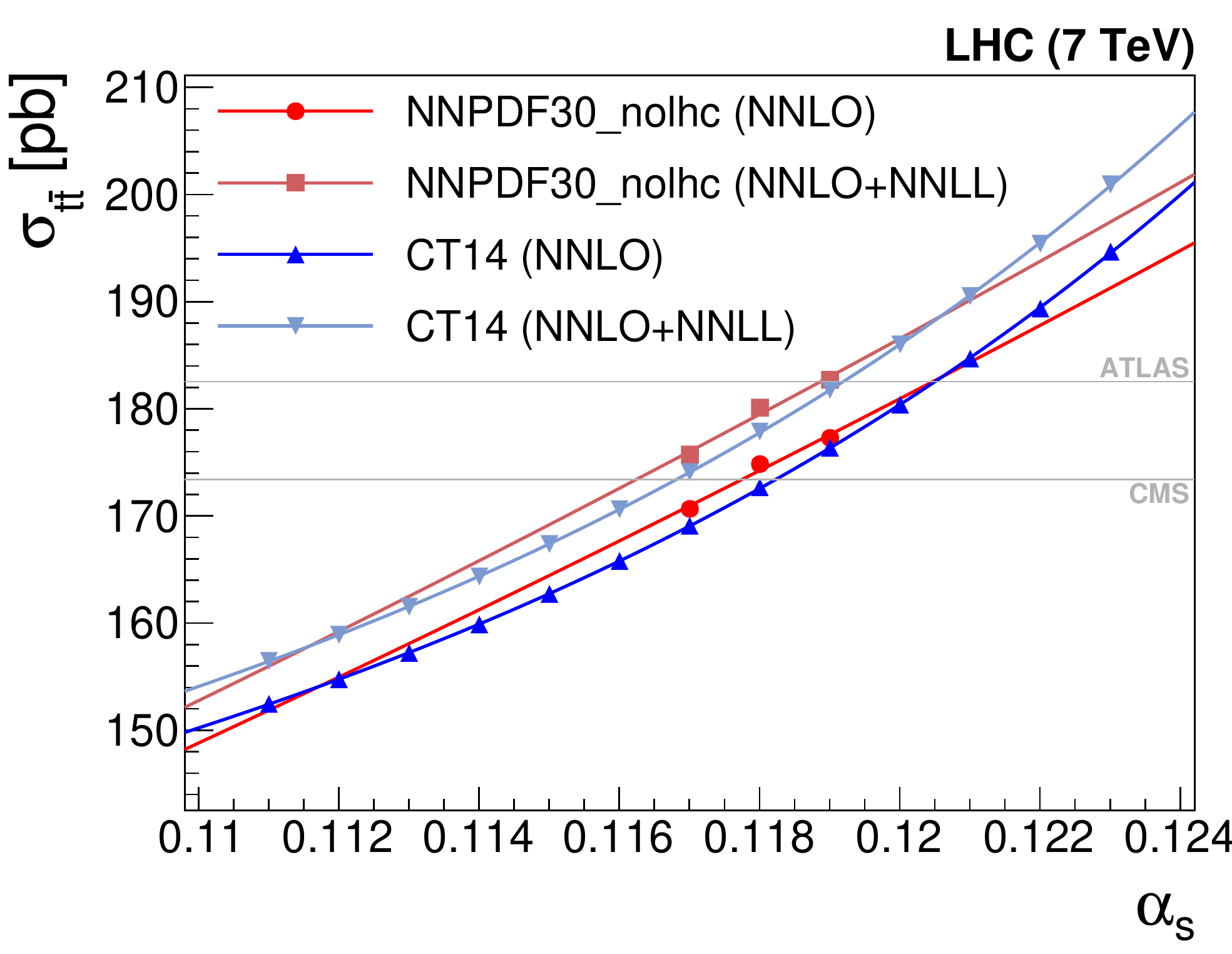}
&
\includegraphics[width=\FitFigureWidth\linewidth]{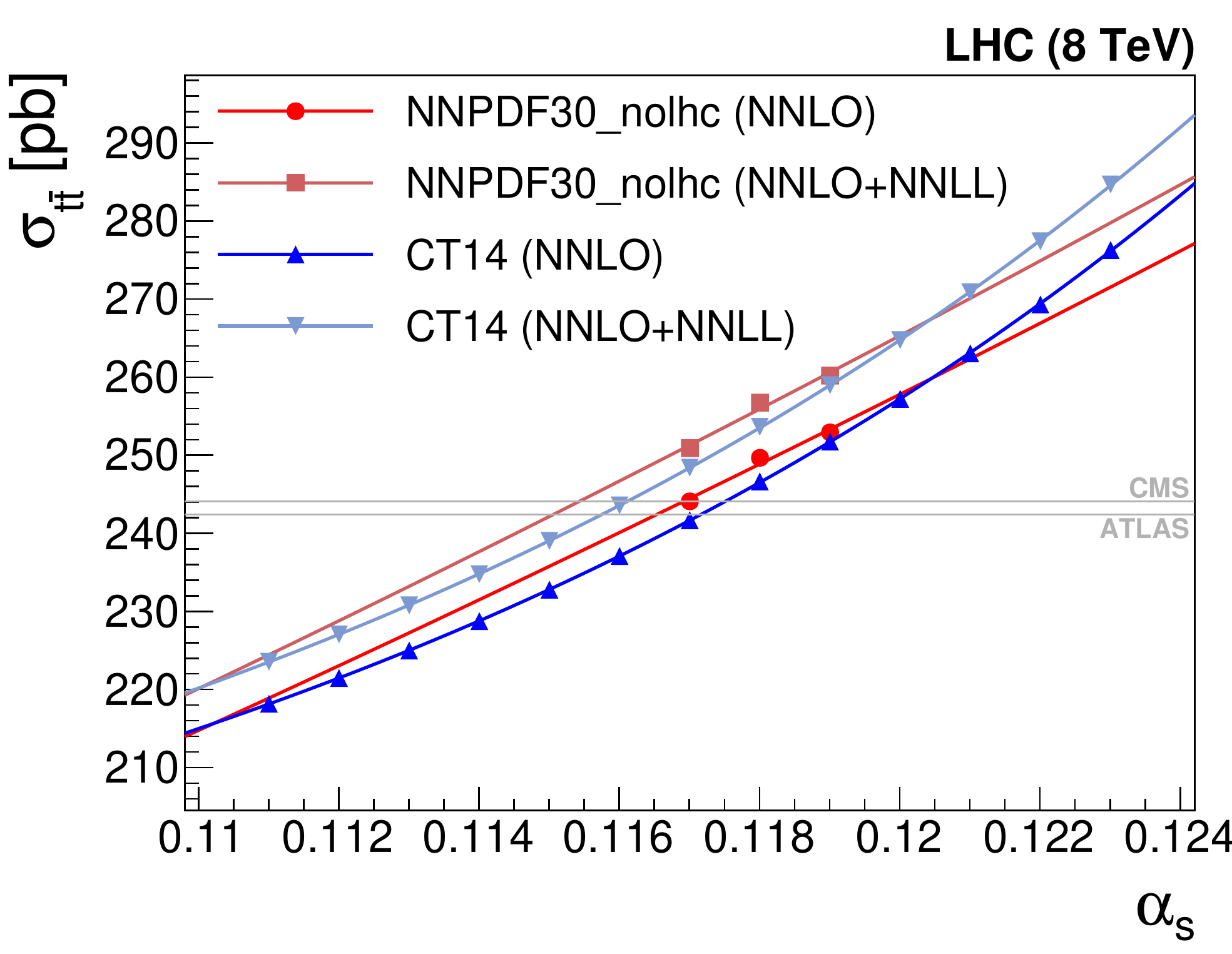}
\\[-8pt]
(a) & (b) \\[-3pt]
\includegraphics[width=\FitFigureWidth\linewidth]{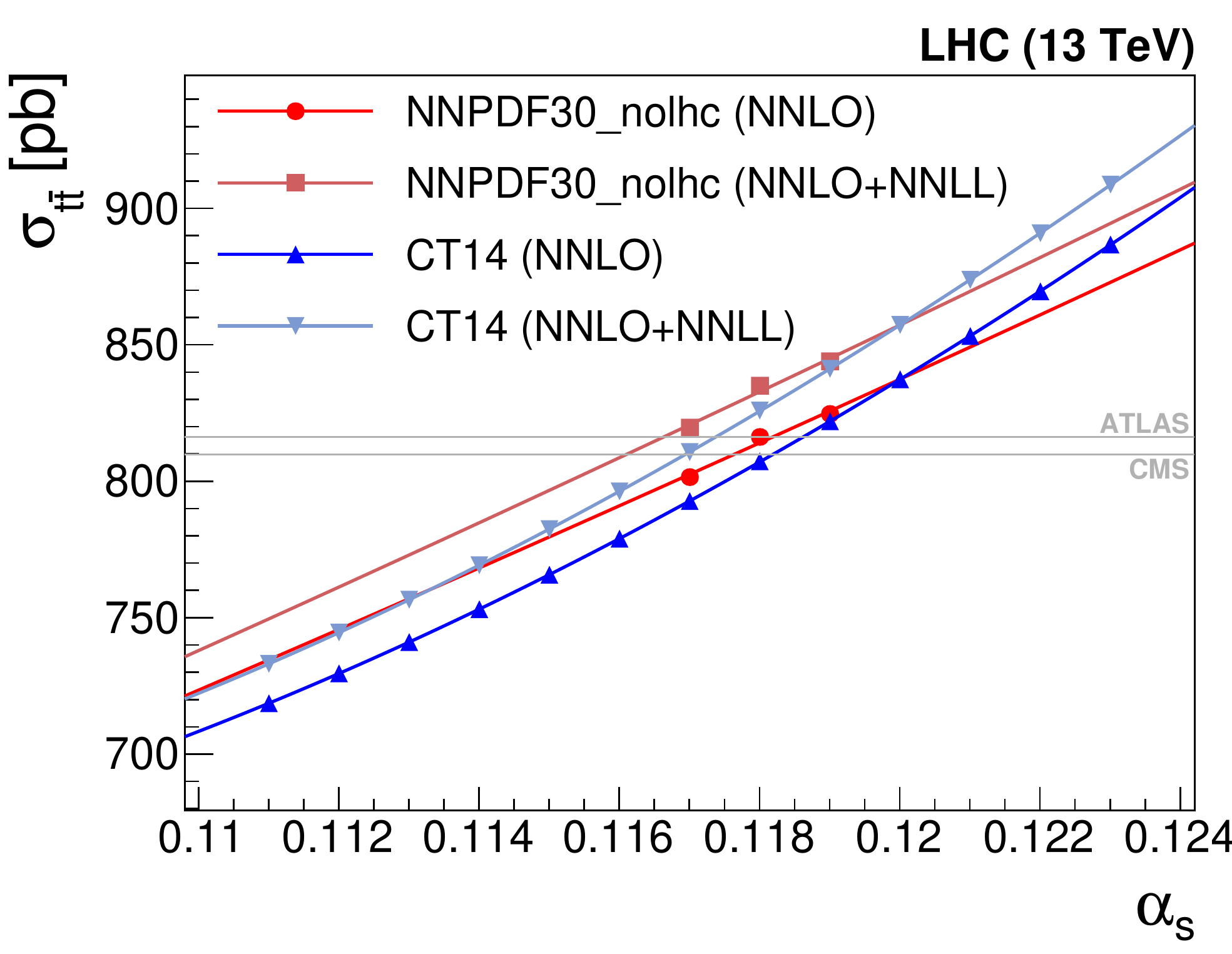}
&
\includegraphics[width=\FitFigureWidth\linewidth]{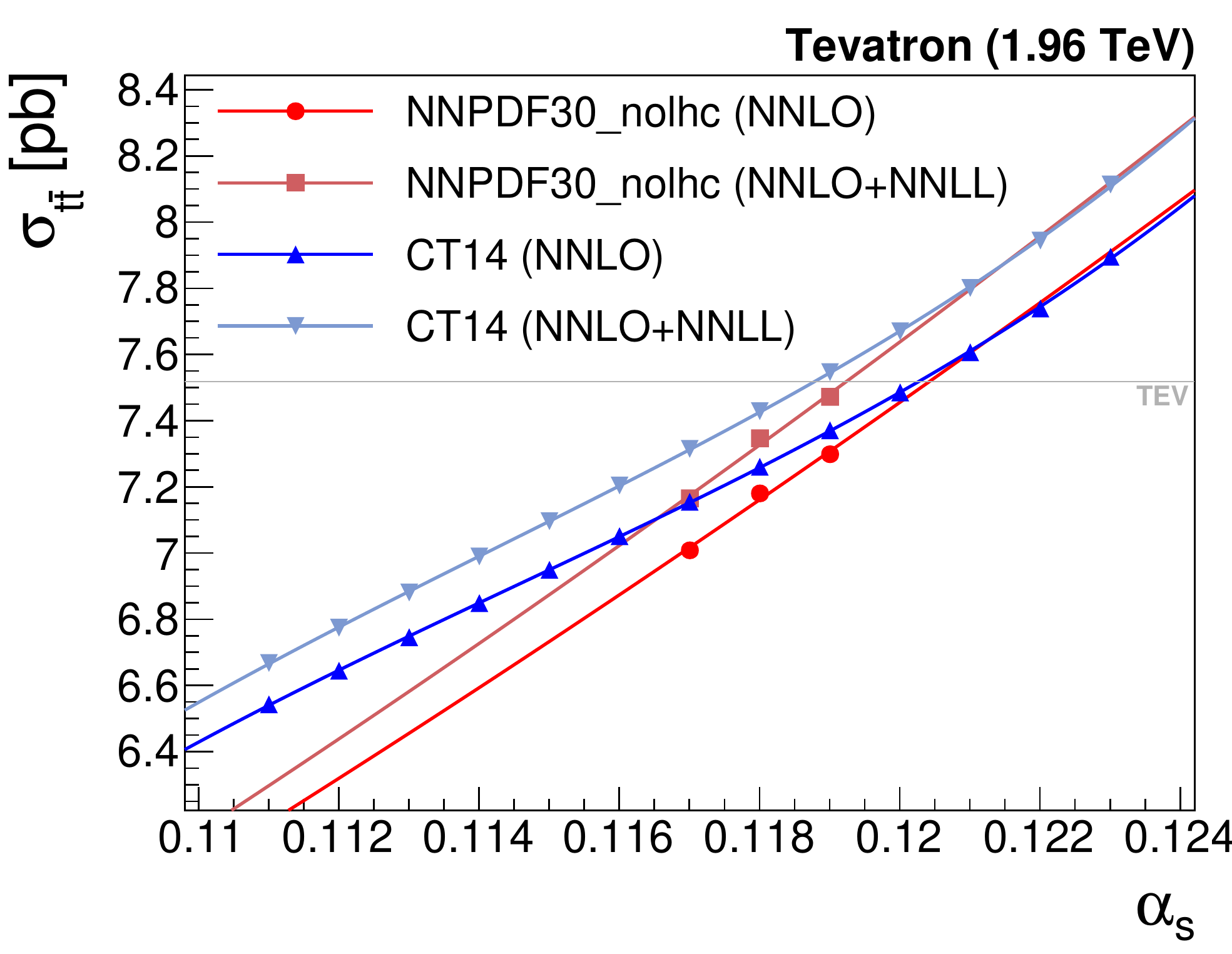}
\\[-8pt]
(c) & (d) \\[6pt]
\end{tabular}
\vspace{-0.3cm}
\caption{
  Predicted cross section as a function of $\as$. The points are the cross
  sections calculated using the \texttt{Top++} program~\cite{topplusplus}, and
  the line is our polynomial fit.
  The plot also includes horizontal lines corresponding to the central
  values of the measured cross sections, adjusted to correspond to the
  same top mass as the theory cross sections ($m_t = m_t^\text{ref} =
  173.2\GeV$), cf.~Section~\ref{sec:top-mass-dependence}.
}
\label{fig:FitsToPrediction}
\end{figure}

We use PDF uncertainties calculated at the 68\% confidence level,
following the error propagation prescription from the individual PDF
groups.
The uncertainties from the CT14 PDF set, which are provided at a 90\%
confidence level by default, are scaled by a factor of
$1/(\sqrt2 \, \text{erf}^{-1}( 0.90) )\simeq 0.608 $.

The predicted cross sections for both PDF sets, with NNLO and
NNLO+NNLL calculations, are listed in
Table~\ref{tab:includedpredictions}.
The cross sections are $1{-}3\%$ higher when including NNLL
contributions.
The scale uncertainties are in the $4{-}6\%$ range for the NNLO
results and get reduced by between one third and one half when
including NNLL terms.
At LHC energies, the cross sections with NNPDF30\_nolhc are about $1\%$ larger
than those with CT14, however the opposite pattern is seen at
Tevatron. 
Finally, the PDF uncertainties are somewhat larger with NNPDF\_nolhc than
with CT14.

To understand the final errors on the $\as$ determination it is
important also to examine how the predicted cross sections depend on
$\as$, a result of the $\as$ dependence both of the hard cross section
and of the PDFs.
This is shown in Fig.~\ref{fig:FitsToPrediction}: points correspond
to the values of $\as$ for which the given PDF is available, and lines
correspond to a fit for $\ln \stt$ using a polynomial of $\ln \as$. 
We use polynomials of degree $3$ and $1$ respectively for the CT14 and
NNPDF30\_nolhc PDFs, chosen based on the available number of $\as$ points and
requirements of stability of the extrapolation beyond the available
$\as$ points.
A steeper slope of the $\as$ dependence (also quoted at $\as=0.118$ in
the last column of Table~\ref{tab:includedpredictions})
leads to a smaller final error on $\as$ for any given source of
uncertainty on $\stt$. 
For LHC energies, CT14 is generally steeper, while at the Tevatron
it is NNPDF\_nolhc that is steeper.
Note also that CT14 curves have substantial curvature, and this will
induce asymmetric uncertainties for $\as$, even in the case of
uncertainties on the cross section that are symmetric.

\subsection{Top-mass dependence}
\label{sec:top-mass-dependence}

The top-quark pole mass is taken to be $173.2 \pm 0.87\,$GeV, which
is consistent with the world average value computed by the Particle Data
Group~\cite{PDG}. 
The experimentally measured cross section, $\stt^{\text{exp}} (\mt)$,
depends on $\mt$ through the acceptance corrections, whose
parametrization is given together with the individual
measurements.
The uncertainty on the experimentally measured cross section due to the top-quark pole mass is given in Tab.~\ref{tab:includedmeasurements}, where the uncertainty was calculated by shifting the top mass up and down by its uncertainty.
An increase in the top mass leads to a decrease in the measured total
cross section.
This is because the experiments effectively measure a fiducial cross
section (which is independent of $m_t$) and then extrapolate it to a
total cross section by dividing by the acceptance for the fiducial
cross section.
For larger values of $m_t$ the acceptance is larger, since decay
products are more likely to pass transverse momentum cuts, and so the resulting
total cross section is lower.
The theoretically predicted cross section,
$\stt^{\text{pred}} ( \mt )$, also depends on $\mt$, because of the
structure of the underlying hard cross section and the $x$-dependence
of the PDFs, cf.\ Tab.~\ref{tab:includedpredictions}.
It too decreases for an increase in the cross section, and this
effect is larger than for the measured cross section.

To define a single error contribution associated with the
top-mass uncertainty, it is convenient to absorb these different
sources of $m_t$ dependence into an effective predicted cross section,
\begin{equation}
  \label{eq:eff-mtop-dep}
\sigma_{t\bar{t}}^{\text{eff}}(m_t) = 
    \stt^{\text{pred}} (  \mt  ) \cdot 
        \frac{
            \stt^{\text{exp}} (  \mt^\rf  )
            }{
            \stt^{\text{exp}} (  \mt  )
            }\,,
\end{equation}
where $m_t^\rf = 173.2$ is the central value of the world average top
mass.
For $m_t = m_t^\rf$, this effective predicted cross section coincides
with the actual predicted one.

The final uncertainty on the effective predicted cross section
associated with the error of $\Delta m_t = 0.87\GeV$ on the world
average top mass is then given by
\begin{equation}
  \label{eq:mtop-uncertainty-impact}
  \sigma_{t\bar{t}}^{\text{eff}}(m_t^\rf \pm \Delta m_t) - 
  \sigma_{t\bar{t}}^{\text{eff}}(m_t^\rf)\,.
\end{equation}
This can be used in our $\as$ determination in a manner similar to
any of the theoretical and PDF uncertainties on the predicted cross
section.
To a good approximation, the final top-mass uncertainty on the
effective cross section is equal to the difference between the
percentage uncertainties in Tabs.~\ref{tab:includedmeasurements} and
\ref{tab:includedpredictions}.
%

\subsection{Strong coupling determination procedure}
\label{sec:determination-procedure}

In the determination of $\as$ from $\stt$, the theory prediction is
treated as a Bayesian prior (one prior for any given value of $\as$)
and the experimental result as a likelihood function.  The
multiplication of these is the joint posterior probability function
from which $\as$ and its uncertainties are determined after
marginalisation of $\stt$.  The procedure is mostly analogous to that
used by the CMS Collaboration in Ref.~\cite{CMS-ttbar-alphas}.

The construction of the Bayesian prior from the theory dependence
necessitates a single probability distribution function given all
individual theory uncertainties. 
The three theory uncertainties are each interpreted as corresponding
to an asymmetric Gaussian function:
\begin{equation}
f^{\text{Unc. source}} \; (\stt \,|\, \as) = 
\left\{
    \begin{array}{ll}
        \frac{1}{\sqrt{2\pi} \Delta_-}
        \, \text{e}  \,^{-\frac12 \left(
        \frac{
            \stt - \stt^{\text{pred}}(\as)
            }{
            \Delta_-
            }
        \right)^2 }
        & \quad \text{if} \; \stt \leq \stt^{\text{pred}}
        \\[20pt]
        \frac{1}{\sqrt{2\pi} \Delta_+}
        \, \text{e}  \,^{ -\frac12\left(
        \frac{
            \stt - \stt^{\text{pred}}(\as)
            }{
            \Delta_+
            }        
        \right)^2 }
        & \quad \text{if} \; \stt > \stt^{\text{pred}}
    \end{array}
    \right.
,
\label{eq:asymgauss}
\end{equation}
where $\stt^{\text{pred}}(\as)$ is the predicted central value at a
given value of $\as$, and $\Delta_{+(-)}$ is the positive (negative)
uncertainty from a given theory uncertainty source. This function has
the advantage that the integral normalizes naturally to one, and that
the integral from $(\stt^{\text{pred}}-\Delta_-)$ to
$(\stt^{\text{pred}}+\Delta_+)$ corresponds to a 68\% 
confidence interval.  
On average there is a 20\% difference between $\Delta_+$ and $\Delta_-$,
and up to a difference of about 85\% for the most asymmetric uncertainty.
The central value for $\stt$ corresponds to the median of the
distribution.

\begin{figure}[htb]
\centering
\begin{tabular}{cc}
\includegraphics[width=0.5\linewidth]{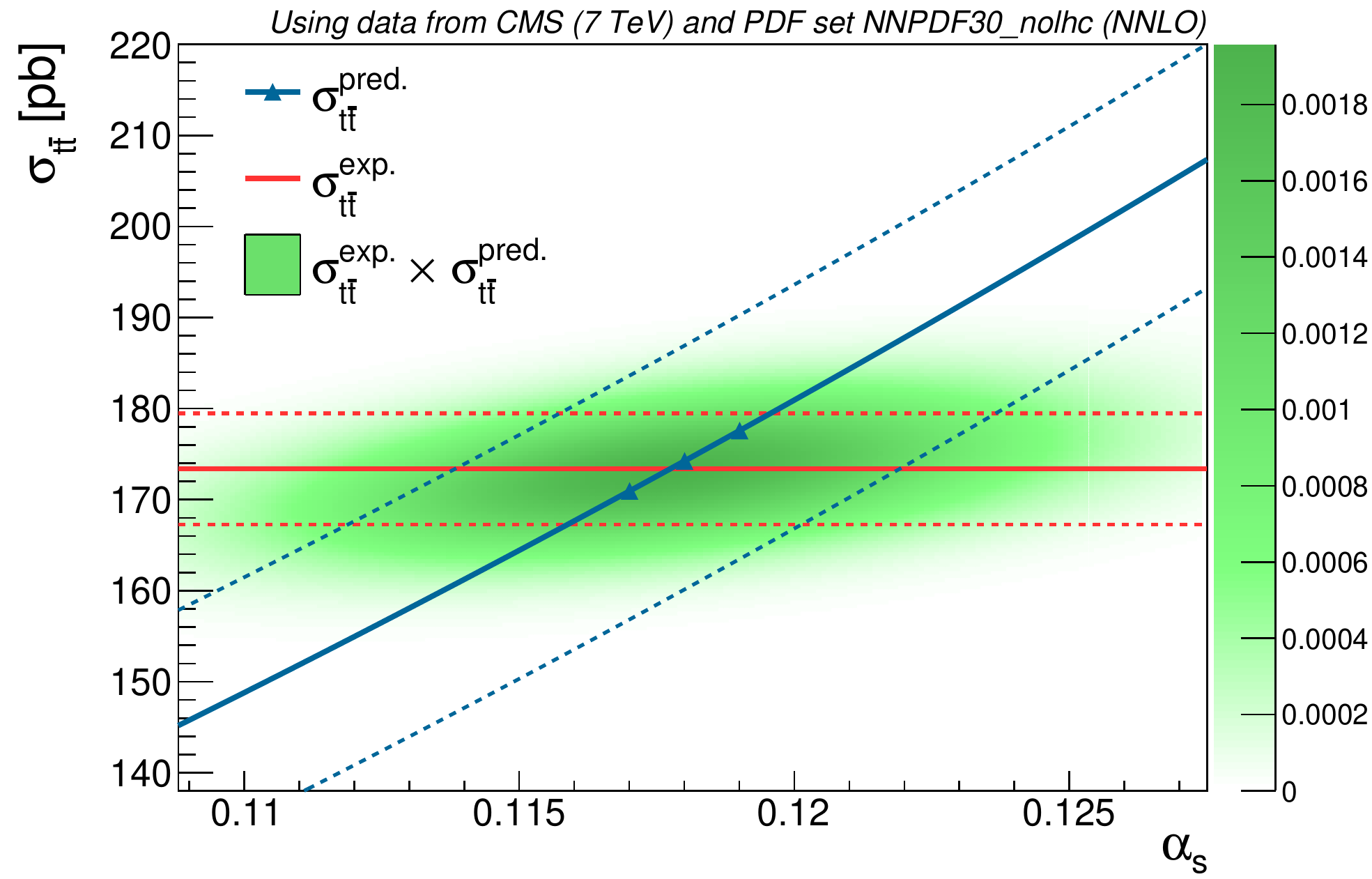}
&
\includegraphics[width=0.5\linewidth]{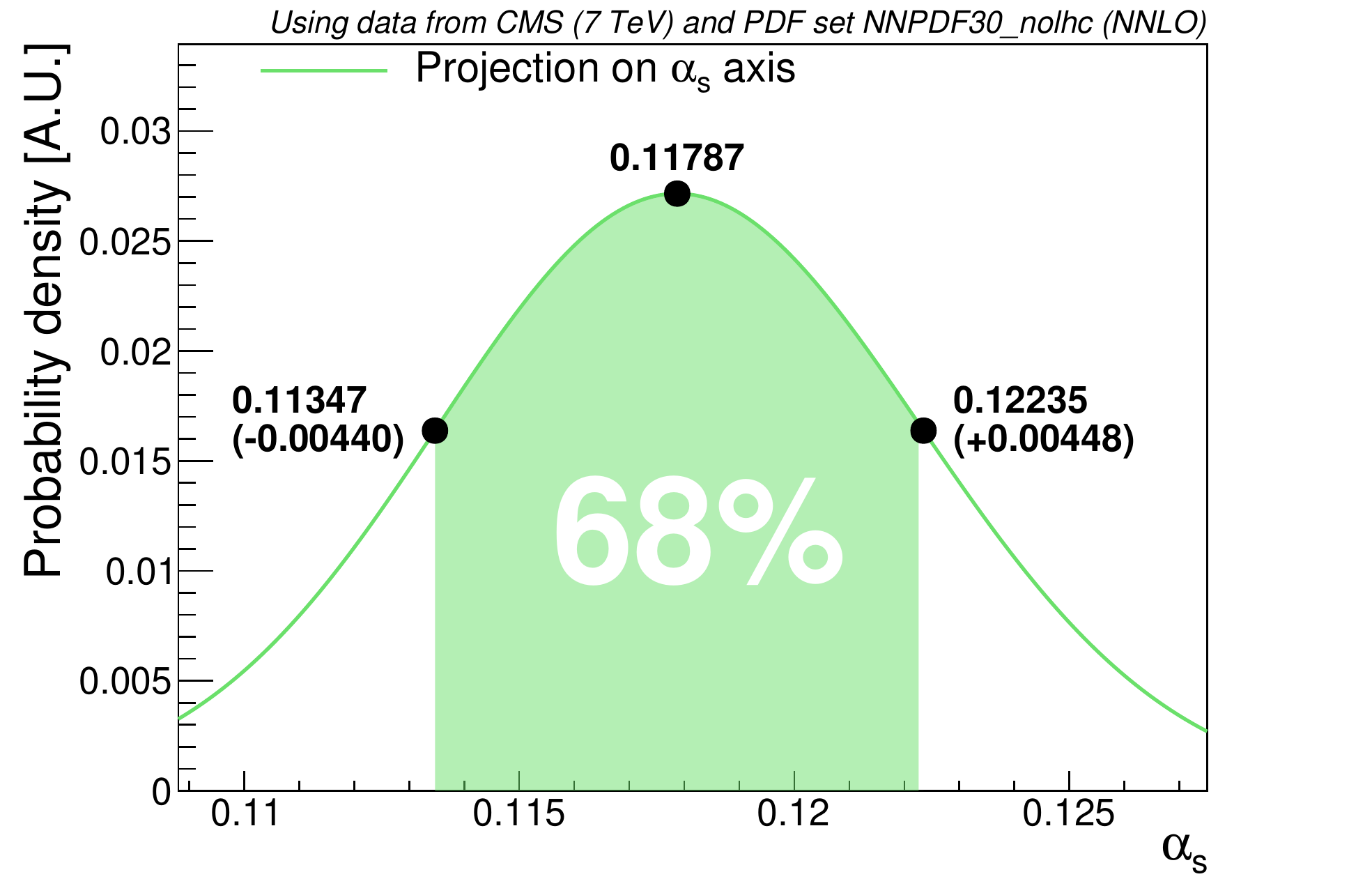}
\\
(a) & (b)
\end{tabular}
\vspace{-0.3cm}
\caption{
  (a) The central values and 1$\sigma$ deviations for the predicted cross
  section ($f^{\text{pred}}(\stt\,|\,\as)$, blue oblique lines) and
  the experimental cross section ($f^{\text{exp}}(\stt\,|\,\as)$,
  red horizontal lines) 
  and the product of the probability distribution functions (green
  shading).
  The markers on the predicted cross
  section indicate the fit points from \texttt{top++2.0}. (b)
  Marginalisation of the joint posterior with Bayesian confidence
  interval.
  }
\label{fig:fullproc}
\end{figure}

The combined probability distribution function of the predicted cross
section, $f^{\text{pred}}(\stt\,|\,\as)$, is computed by taking the
numerical convolution of the individual asymmetric Gaussian functions:
\begin{equation}
f^{\text{pred}}(\stt\,|\,\as) = 
    f^{\text{PDF}}(\stt\,|\,\as)    \otimes
    f^{\mt}(\stt\,|\,\as)           \otimes
    f^{\text{Scale}}(\stt\,|\,\as)\,,
\end{equation}
where the convolution is performed such that the probability distribution
functions are centred around $\stt^\text{pred}$.
%
While the individual uncertainty distributions contain a discontinuity
at $\stt = \stt^{\text{pred}}(\as)$, the convolution is a smooth
function.
The dependence on $\as$ of the width of the uncertainty band is neglected.%
\footnote{ With this approach of fixed absolute uncertainties on $\stt$,
    theory uncertainties on $\as$ will turn out relatively smaller for
    determinations with a higher central $\as$ value.
    One concern is that this might affect the relative weights of
    different determinations in the combination that is described
    later in Sect.~\ref{sec:combination}.
    To address this concern, a cross-check was performed in which the
    individual theory errors from our procedure were scaled relative
    to the default approach by a factor
    $\frac{ \as^\text{determination} }{ \as^\text{ref} }$.
    That is equivalent to taking fixed relative (rather than fixed
    absolute) theory uncertainties on $\stt$.
    With the combination procedure of section~\ref{sec:combination},
    the difference induced by this change was below the per mille
    level.
    For the alternative combination procedure in
    Appendix~\ref{sec:appendix}, the effect is less than half a 
    percent on $\as$, which remains much smaller than the difference
    between the two combination procedures.  }
The probability distribution function of the predicted cross section
is multiplied by the probability distribution function of the measured
cross section $f^{\text{exp}}(\stt\,|\,\as)$, yielding the joint
Bayesian posterior in terms of $\as$ and $\stt$. The Bayesian
confidence interval of $\as$ can be computed through marginalisation
of the posterior by integrating over $\stt$:
\begin{equation}
  \label{eq:likelihood-master-equation}
  L(\as) = \int f^{\text{pred}}(\stt\,|\,\as) 
  \cdot 
  f^{\text{exp}}(\stt\,|\,\as) \; \text{d}\stt\,.
\end{equation}
Here, $f^{\text{exp}}(\stt\,|\,\as)$ is taken to be independent of
$\as$. Technically a small dependence on $\as$ is introduced in
$f^{\text{exp}}(\stt\,|\,\as)$ through the acceptance corrections;
however, in the region of relevance around $\as^{\text{ref}} = 0.118$,
the effect of this on the uncertainty of the cross section is below
the percent level~\cite{CMS-ttbar-alphas}, and can thus be safely
neglected.  
The marginalised joint posterior $L(\as)$ can be treated
as a probability distribution function. 
The central value for the $\as$ determination is taken to be the
location of the peak of $L(\as)$, and the
uncertainty is extracted by computing the 68\% confidence interval
whose left and right bounds are at equal height.\footnote{This is
  somewhat different from the prescription to define an asymmetric
  probability distribution in Eq.~(\ref{eq:asymgauss}), but coincides
  with widespread practice in ATLAS and CMS likelihood fits.}
The procedure is illustrated in Fig.~\ref{fig:fullproc}, showing the
experimental and theory probability distribution functions and the
unmarginalised posterior (Fig.~\ref{fig:fullproc}(a)) as well as the
marginalised posterior with extracted central value and uncertainties
(Fig.~\ref{fig:fullproc}(b)).

The combination of determinations from different experiments
necessitates a breakdown of the total uncertainty into components that
can be assigned to the individual uncertainty sources.  To this end,
the determination is repeated each time omitting a different
uncertainty source, and the squared difference of the resulting
uncertainty with respect to the total uncertainty is computed.  A
relative contribution to the total uncertainty is then computed per
uncertainty source. 

\subsection{Individual results for $\as$ per $\sigma_{t\bar t}$ measurement}
\label{sec:results-by-exp}

The results of our $\as$ determination are listed for the CT14nnlo
PDF set in Tables~\ref{tab:determination_NNLO_CT14} and
\ref{tab:determination_NNLO_NNLL_CT14} and for the
NNPDF30\_nolhc PDF set in
Tables~\ref{tab:determination_NNLO_NNPDF30nolhc} and
\ref{tab:determination_NNLO_NNLL_NNPDF30nolhc}.

The individual $\as$ determinations are all compatible with the world
average to within uncertainties.
The central values are rather similar with the CT14 and NNPDF sets.
The largest individual sources of uncertainty on $\as$ are the PDF
uncertainties and the scale uncertainties.
For the LHC determinations, the PDF uncertainties tend to be larger
with NNPDF, in part a consequence of the larger uncertainties in the
cross section in Table~\ref{tab:includedpredictions}. 
However the other uncertainties are also larger with NNPDF, because
of its weaker dependence on $\as$.

The NNLO+NNLL determinations all have smaller $\as$ results,
consistent with the larger cross sections in
Table~\ref{tab:includedpredictions}.
The scale uncertainties are also noticeably smaller.
Other uncertainties are largely unchanged.

A final comment concerns the somewhat larger scale, $m_t$ and PDF
uncertainties with the CT14 PDF for the CMS 7 TeV case as compared to
the ATLAS 7 TeV case, or also ATLAS 8 TeV as compared to ATLAS 7 TeV.
In general with the CT14 PDF, a smaller value of $\as$ corresponds to
larger uncertainties, because the $\as$ dependence of the cross
section is weaker for small $\as$ values,
cf.~Fig.~\ref{fig:FitsToPrediction}. 
Note however, that the scale and other uncertainties on the cross
section predictions have been evaluated only for the reference value
of $\as=0.118$, and in general the question of how one should
correlate uncertainties with the central value is a delicate
one.\footnote{As an example, imagine that we had used scale
  uncertainties that depended on $\as$: then for an experimental
  measurement with cross section that fluctuates low, one would deduce
  a smaller scale uncertainty than for a cross section that fluctuates
  high; when combining them, depending on the procedure, this might
  then lead to a larger weight for the smaller value of $\as$.}
Accordingly one should be wary of reading too much into the variation
of uncertainties with the central $\as$ value.

\newcommand{\ErrTableWidth}{0.8cm}
\begin{table*}[ht] 
{\scriptsize  
\renewcommand{\arraystretch}{1.4}
\begin{center} 
\begin{tabular}{l c c c c c c c c c }
\toprule
&
\cell{\ErrTableWidth}{Center} & 
\cell{\ErrTableWidth}{Stat.} & 
\cell{\ErrTableWidth}{Syst.} & 
\cell{\ErrTableWidth}{Lumi.} & 
\cell{\ErrTableWidth}{$E_{\text{beam}}$} & 
\cell{\ErrTableWidth}{PDF} & 
\cell{\ErrTableWidth}{Scale} & 
\cell{\ErrTableWidth}{$\mt$} & 
\cell{\ErrTableWidth}{Total} \\ 
\midrule
ATLAS (7\,TeV)        & 0.1205 & ${}_{-0.0009}^{+0.0007}$ & ${}_{-0.0012}^{+0.0009}$ & ${}_{-0.0010}^{+0.0008}$ & ${}_{-0.0001}^{+0.0001}$ & ${}_{-0.0021}^{+0.0015}$ & ${}_{-0.0021}^{+0.0021}$ & ${}_{-0.0012}^{+0.0009}$ & ${}_{-0.0036}^{+0.0030}$ \\
ATLAS (8\,TeV)        & 0.1171 & ${}_{-0.0004}^{+0.0003}$ & ${}_{-0.0014}^{+0.0011}$ & ${}_{-0.0013}^{+0.0010}$ & ${}_{-0.0002}^{+0.0001}$ & ${}_{-0.0025}^{+0.0017}$ & ${}_{-0.0026}^{+0.0027}$ & ${}_{-0.0015}^{+0.0011}$ & ${}_{-0.0044}^{+0.0037}$ \\
ATLAS (13\,TeV)       & 0.1187 & ${}_{-0.0006}^{+0.0006}$ & ${}_{-0.0021}^{+0.0017}$ & ${}_{-0.0014}^{+0.0012}$ & ${}_{-0.0001}^{+0.0001}$ & ${}_{-0.0016}^{+0.0014}$ & ${}_{-0.0024}^{+0.0026}$ & ${}_{-0.0013}^{+0.0011}$ & ${}_{-0.0041}^{+0.0038}$ \\
CMS (7\,TeV)          & 0.1182 & ${}_{-0.0007}^{+0.0005}$ & ${}_{-0.0014}^{+0.0010}$ & ${}_{-0.0013}^{+0.0009}$ & ${}_{-0.0002}^{+0.0001}$ & ${}_{-0.0025}^{+0.0017}$ & ${}_{-0.0025}^{+0.0025}$ & ${}_{-0.0014}^{+0.0010}$ & ${}_{-0.0043}^{+0.0035}$ \\
CMS (8\,TeV)          & 0.1175 & ${}_{-0.0004}^{+0.0003}$ & ${}_{-0.0015}^{+0.0011}$ & ${}_{-0.0016}^{+0.0012}$ & ${}_{-0.0001}^{+0.0001}$ & ${}_{-0.0024}^{+0.0017}$ & ${}_{-0.0026}^{+0.0026}$ & ${}_{-0.0014}^{+0.0010}$ & ${}_{-0.0044}^{+0.0037}$ \\
CMS (13\,TeV)         & 0.1183 & ${}_{-0.0007}^{+0.0006}$ & ${}_{-0.0030}^{+0.0025}$ & ${}_{-0.0015}^{+0.0013}$ & ${}_{-0.0001}^{+0.0002}$ & ${}_{-0.0017}^{+0.0014}$ & ${}_{-0.0025}^{+0.0026}$ & ${}_{-0.0010}^{+0.0009}$ & ${}_{-0.0047}^{+0.0042}$ \\
Tevatron (1.96\,TeV)  & 0.1202 & ${}_{-0.0018}^{+0.0013}$ & ${}_{-0.0026}^{+0.0019}$ & ${}_{-0.0019}^{+0.0014}$ & ${}_{-0.0000}^{+0.0000}$ & ${}_{-0.0020}^{+0.0014}$ & ${}_{-0.0027}^{+0.0024}$ & ${}_{-0.0009}^{+0.0006}$ & ${}_{-0.0050}^{+0.0039}$ \\

\bottomrule
\end{tabular} 
\end{center} 
\caption{
    \small Results for the strong coupling evaluated at the Z-boson mass scale
    and individual uncertainty contributions. 
    These are based on cross sections calculated at NNLO
    using the CT14nnlo series of PDFs.
    \label{tab:determination_NNLO_CT14}
    }
} 
\end{table*} 

\begin{table*}[ht] 
{\scriptsize  
\renewcommand{\arraystretch}{1.4}
\begin{center} 
\begin{tabular}{l c c c c c c c c c }
\toprule
&
\cell{\ErrTableWidth}{Center} & 
\cell{\ErrTableWidth}{Stat.} & 
\cell{\ErrTableWidth}{Syst.} & 
\cell{\ErrTableWidth}{Lumi.} & 
\cell{\ErrTableWidth}{$E_{\text{beam}}$} & 
\cell{\ErrTableWidth}{PDF} & 
\cell{\ErrTableWidth}{Scale} & 
\cell{\ErrTableWidth}{$\mt$} & 
\cell{\ErrTableWidth}{Total} \\ 
\midrule
ATLAS (7\,TeV)        & 0.1206 & ${}_{-0.0009}^{+0.0009}$ & ${}_{-0.0013}^{+0.0012}$ & ${}_{-0.0011}^{+0.0010}$ & ${}_{-0.0001}^{+0.0002}$ & ${}_{-0.0027}^{+0.0025}$ & ${}_{-0.0025}^{+0.0029}$ & ${}_{-0.0013}^{+0.0012}$ & ${}_{-0.0043}^{+0.0044}$ \\
ATLAS (8\,TeV)        & 0.1166 & ${}_{-0.0004}^{+0.0004}$ & ${}_{-0.0013}^{+0.0012}$ & ${}_{-0.0012}^{+0.0011}$ & ${}_{-0.0002}^{+0.0001}$ & ${}_{-0.0026}^{+0.0024}$ & ${}_{-0.0026}^{+0.0032}$ & ${}_{-0.0014}^{+0.0013}$ & ${}_{-0.0043}^{+0.0045}$ \\
ATLAS (13\,TeV)       & 0.1183 & ${}_{-0.0007}^{+0.0007}$ & ${}_{-0.0024}^{+0.0022}$ & ${}_{-0.0017}^{+0.0016}$ & ${}_{-0.0001}^{+0.0002}$ & ${}_{-0.0021}^{+0.0020}$ & ${}_{-0.0029}^{+0.0035}$ & ${}_{-0.0015}^{+0.0015}$ & ${}_{-0.0049}^{+0.0051}$ \\
CMS (7\,TeV)          & 0.1179 & ${}_{-0.0007}^{+0.0006}$ & ${}_{-0.0013}^{+0.0013}$ & ${}_{-0.0012}^{+0.0011}$ & ${}_{-0.0001}^{+0.0001}$ & ${}_{-0.0028}^{+0.0025}$ & ${}_{-0.0025}^{+0.0030}$ & ${}_{-0.0013}^{+0.0012}$ & ${}_{-0.0044}^{+0.0045}$ \\
CMS (8\,TeV)          & 0.1170 & ${}_{-0.0003}^{+0.0003}$ & ${}_{-0.0014}^{+0.0013}$ & ${}_{-0.0015}^{+0.0014}$ & ${}_{-0.0002}^{+0.0001}$ & ${}_{-0.0026}^{+0.0024}$ & ${}_{-0.0026}^{+0.0032}$ & ${}_{-0.0013}^{+0.0012}$ & ${}_{-0.0044}^{+0.0046}$ \\
CMS (13\,TeV)         & 0.1178 & ${}_{-0.0008}^{+0.0008}$ & ${}_{-0.0034}^{+0.0032}$ & ${}_{-0.0017}^{+0.0016}$ & ${}_{-0.0002}^{+0.0003}$ & ${}_{-0.0021}^{+0.0020}$ & ${}_{-0.0029}^{+0.0034}$ & ${}_{-0.0011}^{+0.0011}$ & ${}_{-0.0054}^{+0.0055}$ \\
Tevatron (1.96\,TeV)  & 0.1205 & ${}_{-0.0014}^{+0.0013}$ & ${}_{-0.0020}^{+0.0019}$ & ${}_{-0.0014}^{+0.0014}$ & ${}_{-0.0000}^{+0.0000}$ & ${}_{-0.0017}^{+0.0015}$ & ${}_{-0.0021}^{+0.0023}$ & ${}_{-0.0007}^{+0.0007}$ & ${}_{-0.0040}^{+0.0039}$ \\

\bottomrule
\end{tabular} 
\end{center} 
\caption{\small As in Table.~\ref{tab:determination_NNLO_CT14}, but
  now using NNLO cross sections with the NNPDF30\_nolhc series of PDFs.} 
\label{tab:determination_NNLO_NNPDF30nolhc}
} 
\end{table*} 

\begin{table*}[ht] 
{\scriptsize  
\renewcommand{\arraystretch}{1.4}
\begin{center} 
\begin{tabular}{l c c c c c c c c c }
\toprule
&
\cell{\ErrTableWidth}{Center} & 
\cell{\ErrTableWidth}{Stat.} & 
\cell{\ErrTableWidth}{Syst.} & 
\cell{\ErrTableWidth}{Lumi.} & 
\cell{\ErrTableWidth}{$E_{\text{beam}}$} & 
\cell{\ErrTableWidth}{PDF} & 
\cell{\ErrTableWidth}{Scale} & 
\cell{\ErrTableWidth}{$\mt$} & 
\cell{\ErrTableWidth}{Total} \\ 
\midrule
ATLAS (7\,TeV)        & 0.1192 & ${}_{-0.0009}^{+0.0007}$ & ${}_{-0.0012}^{+0.0010}$ & ${}_{-0.0010}^{+0.0008}$ & ${}_{-0.0001}^{+0.0001}$ & ${}_{-0.0021}^{+0.0016}$ & ${}_{-0.0014}^{+0.0012}$ & ${}_{-0.0012}^{+0.0010}$ & ${}_{-0.0033}^{+0.0027}$ \\
ATLAS (8\,TeV)        & 0.1158 & ${}_{-0.0004}^{+0.0004}$ & ${}_{-0.0014}^{+0.0011}$ & ${}_{-0.0013}^{+0.0011}$ & ${}_{-0.0002}^{+0.0001}$ & ${}_{-0.0025}^{+0.0019}$ & ${}_{-0.0018}^{+0.0016}$ & ${}_{-0.0015}^{+0.0012}$ & ${}_{-0.0040}^{+0.0032}$ \\
ATLAS (13\,TeV)       & 0.1175 & ${}_{-0.0006}^{+0.0005}$ & ${}_{-0.0020}^{+0.0018}$ & ${}_{-0.0014}^{+0.0012}$ & ${}_{-0.0001}^{+0.0001}$ & ${}_{-0.0016}^{+0.0014}$ & ${}_{-0.0017}^{+0.0017}$ & ${}_{-0.0013}^{+0.0012}$ & ${}_{-0.0037}^{+0.0033}$ \\
CMS (7\,TeV)          & 0.1168 & ${}_{-0.0007}^{+0.0006}$ & ${}_{-0.0015}^{+0.0011}$ & ${}_{-0.0013}^{+0.0010}$ & ${}_{-0.0002}^{+0.0001}$ & ${}_{-0.0026}^{+0.0019}$ & ${}_{-0.0017}^{+0.0014}$ & ${}_{-0.0015}^{+0.0012}$ & ${}_{-0.0040}^{+0.0031}$ \\
CMS (8\,TeV)          & 0.1162 & ${}_{-0.0004}^{+0.0003}$ & ${}_{-0.0015}^{+0.0012}$ & ${}_{-0.0016}^{+0.0013}$ & ${}_{-0.0002}^{+0.0001}$ & ${}_{-0.0024}^{+0.0018}$ & ${}_{-0.0018}^{+0.0016}$ & ${}_{-0.0014}^{+0.0011}$ & ${}_{-0.0040}^{+0.0032}$ \\
CMS (13\,TeV)         & 0.1171 & ${}_{-0.0007}^{+0.0006}$ & ${}_{-0.0029}^{+0.0025}$ & ${}_{-0.0015}^{+0.0013}$ & ${}_{-0.0002}^{+0.0001}$ & ${}_{-0.0017}^{+0.0015}$ & ${}_{-0.0018}^{+0.0017}$ & ${}_{-0.0011}^{+0.0009}$ & ${}_{-0.0043}^{+0.0038}$ \\
Tevatron (1.96\,TeV)  & 0.1188 & ${}_{-0.0017}^{+0.0014}$ & ${}_{-0.0025}^{+0.0021}$ & ${}_{-0.0018}^{+0.0015}$ & ${}_{-0.0000}^{+0.0000}$ & ${}_{-0.0020}^{+0.0015}$ & ${}_{-0.0013}^{+0.0011}$ & ${}_{-0.0009}^{+0.0007}$ & ${}_{-0.0043}^{+0.0035}$ \\

\bottomrule
\end{tabular} 
\end{center} 
\caption{\small As in Table.~\ref{tab:determination_NNLO_CT14}, but
  now using NNLO+NNLL cross sections with the CT14nnlo series of PDFs.}
\label{tab:determination_NNLO_NNLL_CT14}
} 
\end{table*} 

\begin{table*}[ht] 
{\scriptsize  
\renewcommand{\arraystretch}{1.4}
\begin{center} 
\begin{tabular}{l c c c c c c c c c }
\toprule
& 
\cell{\ErrTableWidth}{Center} & 
\cell{\ErrTableWidth}{Stat.} & 
\cell{\ErrTableWidth}{Syst.} & 
\cell{\ErrTableWidth}{Lumi.} & 
\cell{\ErrTableWidth}{$E_{\text{beam}}$} & 
\cell{\ErrTableWidth}{PDF} & 
\cell{\ErrTableWidth}{Scale} & 
\cell{\ErrTableWidth}{$\mt$} & 
\cell{\ErrTableWidth}{Total} \\ 
\midrule
ATLAS (7\,TeV)        & 0.1190 & ${}_{-0.0009}^{+0.0009}$ & ${}_{-0.0012}^{+0.0012}$ & ${}_{-0.0011}^{+0.0010}$ & ${}_{-0.0001}^{+0.0001}$ & ${}_{-0.0026}^{+0.0025}$ & ${}_{-0.0015}^{+0.0016}$ & ${}_{-0.0013}^{+0.0012}$ & ${}_{-0.0037}^{+0.0036}$ \\
ATLAS (8\,TeV)        & 0.1152 & ${}_{-0.0004}^{+0.0004}$ & ${}_{-0.0013}^{+0.0012}$ & ${}_{-0.0012}^{+0.0011}$ & ${}_{-0.0001}^{+0.0001}$ & ${}_{-0.0025}^{+0.0024}$ & ${}_{-0.0017}^{+0.0018}$ & ${}_{-0.0014}^{+0.0013}$ & ${}_{-0.0037}^{+0.0037}$ \\
ATLAS (13\,TeV)       & 0.1168 & ${}_{-0.0007}^{+0.0007}$ & ${}_{-0.0023}^{+0.0022}$ & ${}_{-0.0016}^{+0.0015}$ & ${}_{-0.0002}^{+0.0002}$ & ${}_{-0.0020}^{+0.0019}$ & ${}_{-0.0020}^{+0.0022}$ & ${}_{-0.0015}^{+0.0015}$ & ${}_{-0.0043}^{+0.0043}$ \\
CMS (7\,TeV)          & 0.1163 & ${}_{-0.0006}^{+0.0006}$ & ${}_{-0.0013}^{+0.0012}$ & ${}_{-0.0011}^{+0.0011}$ & ${}_{-0.0001}^{+0.0001}$ & ${}_{-0.0027}^{+0.0026}$ & ${}_{-0.0016}^{+0.0016}$ & ${}_{-0.0013}^{+0.0012}$ & ${}_{-0.0038}^{+0.0037}$ \\
CMS (8\,TeV)          & 0.1155 & ${}_{-0.0003}^{+0.0003}$ & ${}_{-0.0013}^{+0.0013}$ & ${}_{-0.0014}^{+0.0014}$ & ${}_{-0.0001}^{+0.0001}$ & ${}_{-0.0025}^{+0.0024}$ & ${}_{-0.0017}^{+0.0017}$ & ${}_{-0.0013}^{+0.0012}$ & ${}_{-0.0038}^{+0.0037}$ \\
CMS (13\,TeV)         & 0.1163 & ${}_{-0.0007}^{+0.0008}$ & ${}_{-0.0032}^{+0.0031}$ & ${}_{-0.0016}^{+0.0015}$ & ${}_{-0.0002}^{+0.0002}$ & ${}_{-0.0020}^{+0.0019}$ & ${}_{-0.0020}^{+0.0022}$ & ${}_{-0.0011}^{+0.0011}$ & ${}_{-0.0048}^{+0.0047}$ \\
Tevatron (1.96\,TeV)  & 0.1194 & ${}_{-0.0013}^{+0.0013}$ & ${}_{-0.0019}^{+0.0018}$ & ${}_{-0.0014}^{+0.0013}$ & ${}_{-0.0000}^{+0.0000}$ & ${}_{-0.0017}^{+0.0016}$ & ${}_{-0.0010}^{+0.0010}$ & ${}_{-0.0007}^{+0.0007}$ & ${}_{-0.0034}^{+0.0033}$ \\

\bottomrule
\end{tabular} 
\end{center} 
\caption{\small As in Table.~\ref{tab:determination_NNLO_CT14}, but
  now using NNLO+NNLL cross sections with the NNPDF30\_nolhc series of PDFs.} 
\label{tab:determination_NNLO_NNLL_NNPDF30nolhc}
} 
\end{table*}

\section{Combination of $\as$ determinations}
\label{sec:combination}

\subsection{Correlation coefficients}
\label{sec:correlationcoefficients}

A combination of measurements can strongly depend on the assumed or calculated correlations~\cite{BLUE1}.
It is therefore necessary to carefully evaluate the correlation coefficients used for the combination.
In the case of $\as$ determinations many correlations can be reasonably motivated or computed.
The correlation coefficients between individual measurements are motivated per uncertainty source.

\newcommand{\lowersub}[1]{\raisebox{-3pt}{\scriptsize #1}}
\newcommand{\highersub}[1]{\raisebox{2pt}{\scriptsize #1}}

\begin{enumerate}
\item \textit{Statistical uncertainties} are considered uncorrelated
  for all experimental inputs.

\item \textit{Systematic uncertainties} are considered fully
  correlated only for measurements obtained with the same
  detector. This concerns the measurements performed by CMS and ATLAS
  at different centre-of-mass energies.

\item \textit{Uncertainties due to beam energy} are fully correlated
  between ATLAS and CMS and are taken to be correlated across
  energies.
  The beam-energy uncertainty at the Tevatron was tiny and is
  neglected, as outlined in the caption of
  Table~\ref{tab:includedmeasurements}. 

\item
    \textit{Uncertainties due to luminosity} are partially correlated between
    ATLAS and CMS. The correlated component of the luminosity uncertainty
    stems from the uncertainty on the bunch current density and
    similarities in the Van der Meer scan fit model.

    The correlated and uncorrelated uncertainties are estimated using the same
    principles as used for the top-quark pair production cross section combinations
    between ATLAS and CMS at 7 and
    8\,TeV~\cite{topcombination_7TeV,topcombination_8TeV}, updated
    with the latest luminosity
    determinations~\cite{Aad:2013ucp,Aaboud:2016hhf,CMS:2012rua,CMS:2013gfa,CMS:2016eto}.
    The luminosity uncertainty (as a percentage of the top-quark pair
    production cross section) is displayed in Table~\ref{tab:lumitable}.
    The luminosity uncertainties on $\as$ are taken to have the same correlation
    coefficient.

    \newcommand{\tableWidthLumi}{2.5cm}  
    \begin{table*}[ht]
    { 
    \begin{center}  
    \begin{tabular}{l c c c c}
    \toprule
    $\sqrt{s}$                & Experiment & Corr.  & Uncorr. & Total \\
    \midrule
    \multirow{2}{*}{7\,TeV}   &  ATLAS     &  0.46  &  1.72  &  1.78 \\
                              &  CMS       &  0.46  &  2.13  &  2.17 \\
    \multirow{2}{*}{8\,TeV}   &  ATLAS     &  0.60  &  1.84  &  1.94 \\
                              &  CMS       &  0.68  &  2.50  &  2.59 \\
    \multirow{2}{*}{13\,TeV}  &  ATLAS     &  0.36  &  2.29  &  2.32 \\
                              &  CMS       &  0.36  &  2.31  &  2.34 \\
    \bottomrule
    \end{tabular} 
    \end{center} 
    \caption{\small
    Correlated, uncorrelated and total luminosity uncertainties with
    respect to the top-quark pair production cross section (in
    percentages)~\cite{topcombination_7TeV,topcombination_8TeV,Aad:2013ucp,Aaboud:2016hhf,CMS:2012rua,CMS:2013gfa,CMS:2016eto}.
    }
    \label{tab:lumitable}
    }
    \end{table*} 
\end{enumerate}

The uncertainties on the predicted cross sections (due to the PDF, the
top-quark mass and the renormalisation and factorisation scale) 
are generally strongly correlated. 
The combination result strongly depends on the assumed correlation
structure of these theoretical uncertainties if included in the
combination, which is usually not known precisely in particular for
the scale uncertainty.
We therefore adopt
a different procedure: The individual results are
simultaneously shifted up and down by their respective total theory uncertainties, and the
combination is re-evaluated.
The difference between the upper and
lower bounds and the original combination is taken to be the
(asymmetric) theoretical uncertainty.

The impact of the alternative procedure of including also the theory
uncertainties within a single combination is discussed in
Appendix~\ref{sec:appendix}.

\subsection{Combining correlated measurements: Likelihood-based approach}

In order to combine the individual results, we opted for a
likelihood-based approach~\cite{likelihoodapproach}.\footnote{As a cross-check, we also
  used the \textit{Best Linear Unbiased Estimate} procedure
  (BLUE)~\cite{BLUE1,BLUE2}. This is only suitable for symmetric
  errors and in that case we found essentially identical results.}
In this
approach a global likelihood function is constructed from the
probability distribution functions of individual determinations.  
Let us suppose we have $n_m$ measurements of the top cross section and
associated determinations of $\as$. 
For each determination $i$, $\as{}_{,i}$, we have $n_u$ uncorrelated
error components, each specific to that determination.
The magnitude of the $k^\text{th}$ uncorrelated error for
determination $i$ is labelled $\Delta^k_i$.
We additionally have $n_c$ error components that are correlated across
all determinations.
For each of the correlated components, $j$, we introduce a nuisance
parameter $\theta_j$ that is common across all measurements.
Its impact on measurement $i$ is governed by a coefficient
$\delta_i^j$. 
The full set of $\theta_j$ will be denoted $\bm\theta$.

The likelihood will be composed of a product of probability
distribution functions (pdf)\footnote{
\textit{pdf}, for probability density function, is not to be confused with \textit{PDF}, for parton distribution function.
}.
%
For each nuisance parameter we will have one pdf, a Gaussian
distribution with a standard deviation of one:
\begin{equation}
    \text{pdf}_{\theta_j} = 
    \frac{1}{\sqrt{2\pi}} e^{-\theta_j^2/2}
    \,.
\end{equation}
There will also be a pdf for each combination of measurement $i$ and
associated uncorrelated error $\Delta^k_i$. 
It is given by
\begin{equation}
\text{pdf}_{i,\,k}(\as, \bm{\theta}) =
    \frac{1}{\sqrt{2\pi} \Delta_i^k}
    \, \exp{ \displaystyle\left[
    -\frac{
        (\as {}_{,i} \, + \sum_j \theta_j \cdot \delta_i^j - \as)^2
        }{
        2(\Delta_i^k)^2
        }
    \right] }
    \;.
\end{equation}
To address the issue of errors that are not symmetric, we adopt the
following prescription for the $\Delta_i^k$ and $\delta_i^j$:
\begin{equation}
\Delta_i^k =
    \left\{
    \begin{array}{ll}
        \displaystyle
        \Delta_i^{k,\,-}
        & \quad \text{if} \;\; \as \le \as {}_{,i}
        \\[20pt]
        \displaystyle
        \Delta_i^{k,\,+}
        & \quad \text{if} \;\; \as > \as {}_{,i}
    \end{array}
    \right. 
    \;,
\end{equation}
\begin{equation}
\delta_i^j =
    \left\{
    \begin{array}{ll}
        \displaystyle
        \delta_i^{j,\,-}
        & \quad \text{if} \;\; \as \le \as {}_{,i}
        \\[20pt]
        \displaystyle
        \delta_i^{j,\,+}
        & \quad \text{if} \;\; \as > \as {}_{,i}
    \end{array}
    \right. 
    \;.
\end{equation}
An overview of the values used for $\delta_i^{j,\,\pm}$ and $\Delta_i^{k,\,\pm}$
is given in Appendix~\ref{sec:appendix2}.
The probability distribution function of determination $i$ including
all uncorrelated uncertainties is then constructed by convolution:
\begin{equation}
\text{pdf}_{ \as {}_{,i} }(\as, \bm{\theta}) =
    \text{pdf}_{i,\,1}(\as, \bm{\theta}) \otimes \text{pdf}_{i,\,2}(\as, \bm{\theta})
    \otimes \cdots
    \otimes \text{pdf}_{i,\,n_u}(\as, \bm{\theta})
\end{equation}
where the convolution is performed such that the probability distribution
functions are centred around $\as{}_{,i}$.
%
%
The global likelihood function $L( \as , \bm{\theta} )$ is constructed
by multiplication of the probability distribution functions of the
determinations and the nuisance parameters:
\begin{equation}
L( \as , \bm{\theta} ) = 
    \prod_{i=1}^{n_m} \text{pdf}_{ \as {}_{,i} }(\as, \bm{\theta})
    \, \times \,
    \prod_{j=1}^{n_c} \text{pdf}_{\theta_j}
    \,.
    \label{eq:globallikelihoodfunction}
\end{equation}
In order to complete the formalism of a statistical test the test
statistic $q$ is introduced:
\begin{equation}
q( \as ) = -2 \log \frac{ L( \as ,\, \hat{\bm{\theta}}' ) }{ L( \hat{\alpha}_s ,\, \hat{\bm{\theta}} ) }
\;.
\end{equation}
Here $L$ is maximized for variables that carry a hat and in general
$\hat{\bm{\theta}}'$ will take on different values from
$\hat{\bm{\theta}}$.
The quantity $L( \hat{\alpha}_s ,\, \hat{\bm{\theta}} )$ is therefore
the global maximum likelihood, and the ratio cannot be larger than
one. 
The normalisation is such
that $q$ can be treated as $\chi^2$-distributed with one degree of
freedom.

The test statistic $q$ is scanned over a range of $\as$ values. The
minimum of the scan, by construction at $q=0$, is the maximum
likelihood value for $\as$, and the $1\sigma$ confidence interval is
extracted from the interval between the intersection points of the
scan with $q=1$. Any skewness of the parabola of the scan is due to
the inclusion of asymmetric uncertainties.
Figure~\ref{fig:ScanResults} shows the scan and the corresponding
combination results for each of the PDF sets.

\section{Results and discussion}

The combination procedure is performed for each of the two PDF sets
taken into consideration at NNLO and at NNLO+NNLL separately. The
combination results and their unweighted average are displayed
numerically in Table~\ref{tab:combinationresults}, and graphically in
Fig.~\ref{fig:unweightedaverage}.

There is no unique way to quote a final best estimate of $\as$ based
on the results obtained from the different PDF sets and QCD
calculation choices (NNLO v.\ NNLO+NNLL).
An unbiased approach for combining results from different PDFs, in
line with the PDF4LHC recommendations~\cite{PDF4LHC}, is to average
without applying any further weighting.
In accordance with that approach we take the straight average of the mean
values and the uncertainties of the individual combinations.
This coincides with the procedure for combining $\as$ results from a
single class of observables in Ref.~\cite{PDG}.
The final result is
\begin{equation}
\asmz =
    \alphasResultCenter
    \; {}^{\highersub{$+\alphasResultRightError${}}}_{\lowersub{$-\alphasResultLeftError${} }}
\end{equation}
which can be compared to the result of Ref.~\cite{CMS-ttbar-alphas},
$\as(m_Z) = 0.1151^{+0.0028}_{-0.0027}$.
Our central value is larger mainly because recent measurements of the
cross sections are higher than that used in
Ref.~\cite{CMS-ttbar-alphas}, but also in part because of our choice
to take the average of results from NNLO and NNLO+NNLL cross sections
(a $0.6\%$ increase relative to just NNLO+NNLL).
Our symmetrised uncertainty of $\alphasResultSymmPercentage\%$ is
somewhat increased with respect to that of
Ref.~\cite{CMS-ttbar-alphas}, $2.4\%$ (symmetrised).
The difference in uncertainty is due to several choices.
On one hand we have taken a smaller uncertainty on the top-quark mass,
in line with the PDG determination.
One the other hand, we have been somewhat more conservative in our
treatment of theoretical and PDF uncertainties.
Firstly, the choice of treating the scale uncertainties on $\stt$ as a
$68\%$ confidence interval instead of a (flat) $100\%$ confidence
interval increases the scale uncertainty component by roughly a factor
of $\sqrt{3}$.
Secondly, we have used an average of the uncertainties from NNLO and
NNLO+NNLL cross section determinations, which also yields a larger
uncertainty than using NNLO+NNLL cross section determinations only.
Finally, the PDF sets used for the determination were chosen with
minimization of potential biases in mind, rather than the ones with
smallest uncertainty.

\begin{table*}[ht] 
{\scriptsize  
\renewcommand{\arraystretch}{1.4}
\renewcommand{\ErrTableWidth}{0.6cm}
\begin{center} 
\hspace*{-0.75cm}\begin{tabular}{l c c c c c c c c c }
\toprule
& 
\cell{\ErrTableWidth}{Center} & 
\cell{\ErrTableWidth}{Stat.} & 
\cell{\ErrTableWidth}{Syst.} & 
\cell{\ErrTableWidth}{$E_{\text{beam}}$} & 
\cell{\ErrTableWidth}{Lumi.} & 
\cell{\ErrTableWidth}{$\mt$} & 
\cell{\ErrTableWidth}{PDF} & 
\cell{\ErrTableWidth}{Scale} & 
\cell{\ErrTableWidth}{Total} \\ 
\midrule
CT14 {\tiny (NNLO)}            & $0.1184$ & ${}_{-0.0003}^{+0.0003}$ & ${}_{-0.0007}^{+0.0006}$ & ${}_{-0.0001}^{+0.0001}$ & ${}_{-0.0006}^{+0.0006}$ & ${}_{-0.0014}^{+0.0010}$ & ${}_{-0.0023}^{+0.0016}$ & ${}_{-0.0025}^{+0.0025}$ & ${}_{-0.0038}^{+0.0033}$ \\
NNPDF30\_nolhc {\tiny (NNLO)}  & $0.1182$ & ${}_{-0.0003}^{+0.0003}$ & ${}_{-0.0007}^{+0.0007}$ & ${}_{-0.0000}^{+0.0000}$ & ${}_{-0.0007}^{+0.0007}$ & ${}_{-0.0013}^{+0.0012}$ & ${}_{-0.0025}^{+0.0023}$ & ${}_{-0.0026}^{+0.0031}$ & ${}_{-0.0040}^{+0.0042}$ \\
CT14 {\tiny (NNLO+NNLL)}       & $0.1172$ & ${}_{-0.0003}^{+0.0003}$ & ${}_{-0.0007}^{+0.0007}$ & ${}_{-0.0001}^{+0.0001}$ & ${}_{-0.0007}^{+0.0006}$ & ${}_{-0.0014}^{+0.0011}$ & ${}_{-0.0023}^{+0.0017}$ & ${}_{-0.0017}^{+0.0015}$ & ${}_{-0.0033}^{+0.0027}$ \\
NNPDF30\_nolhc {\tiny (NNLO+NNLL)} & $0.1168$ & ${}_{-0.0003}^{+0.0003}$ & ${}_{-0.0007}^{+0.0006}$ & ${}_{-0.0001}^{+0.0001}$ & ${}_{-0.0007}^{+0.0007}$ & ${}_{-0.0013}^{+0.0012}$ & ${}_{-0.0024}^{+0.0023}$ & ${}_{-0.0017}^{+0.0018}$ & ${}_{-0.0034}^{+0.0033}$ \\
\midrule
Average                        & $0.1177$ & ${}_{+0.0003}^{+0.0003}$ & ${}_{+0.0007}^{+0.0007}$ & ${}_{+0.0001}^{+0.0001}$ & ${}_{+0.0007}^{+0.0006}$ & ${}_{+0.0013}^{+0.0012}$ & ${}_{+0.0024}^{+0.0020}$ & ${}_{+0.0021}^{+0.0022}$ & ${}_{-0.0036}^{+0.0034}$ \\
\bottomrule
\end{tabular} 
\end{center} 
\caption{\small Combination results for all PDF sets taken into consideration, at NNLO and NNLO+NNLL.}
\label{tab:combinationresults}
} 
\end{table*}

\newcommand{\ScanFigureWidth}{0.47}  
\begin{figure}[htb]
\centering
\begin{tabular}{ccc}
\includegraphics[width=\ScanFigureWidth\linewidth]{%
    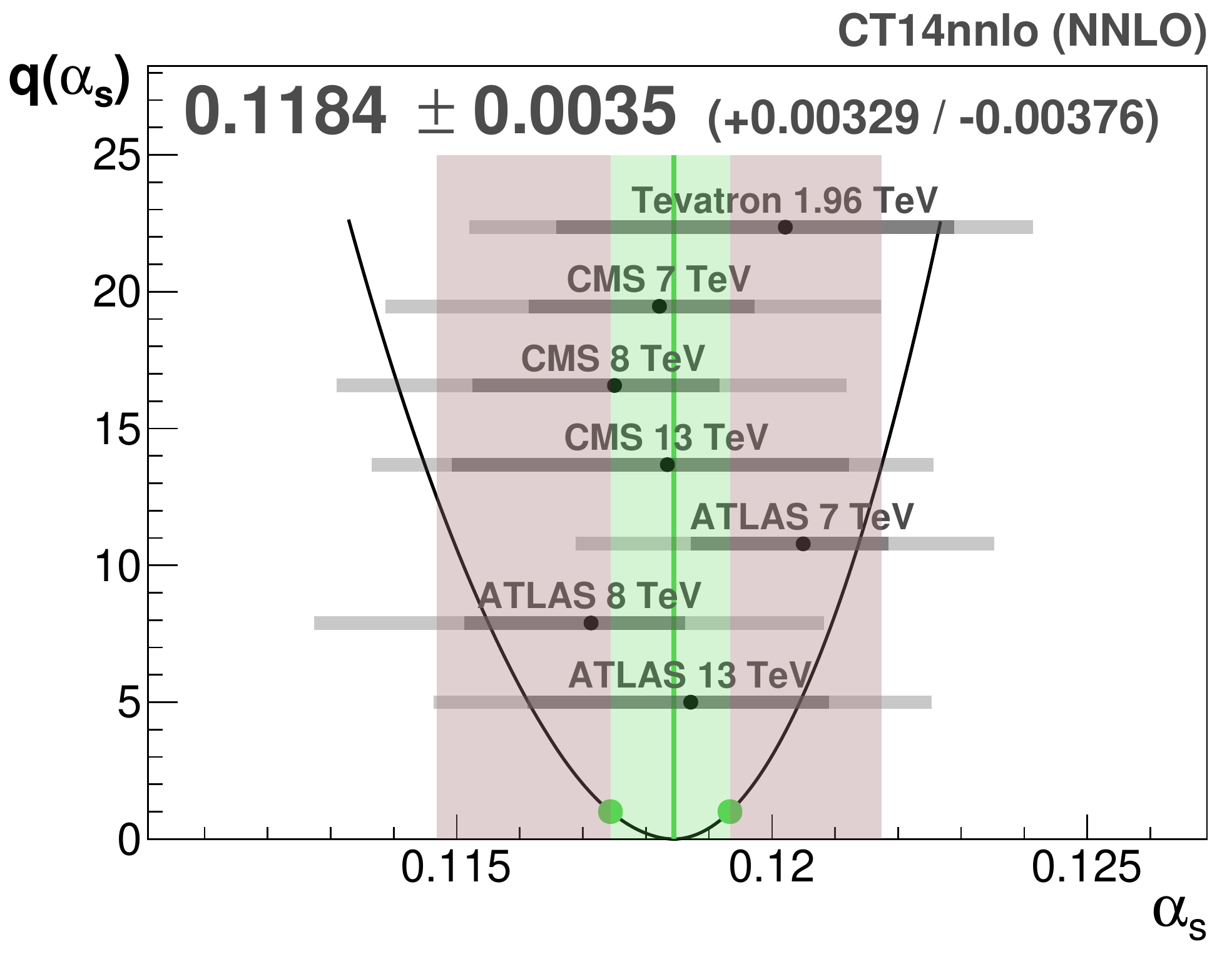%
    }
&
\includegraphics[width=\ScanFigureWidth\linewidth]{%
    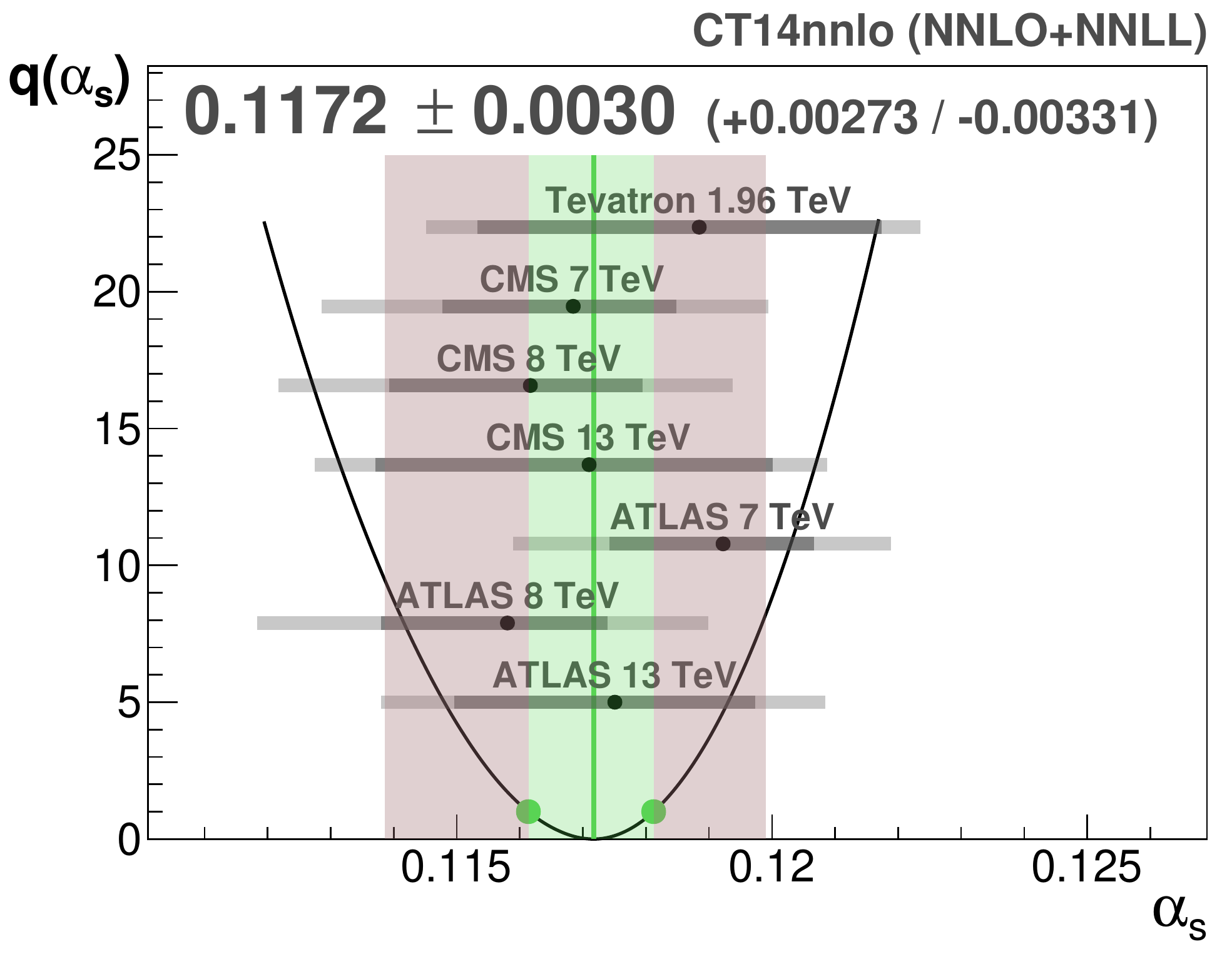%
    }
\\[-6pt]
(a) & (b) \\[8pt]
\includegraphics[width=\ScanFigureWidth\linewidth]{%
    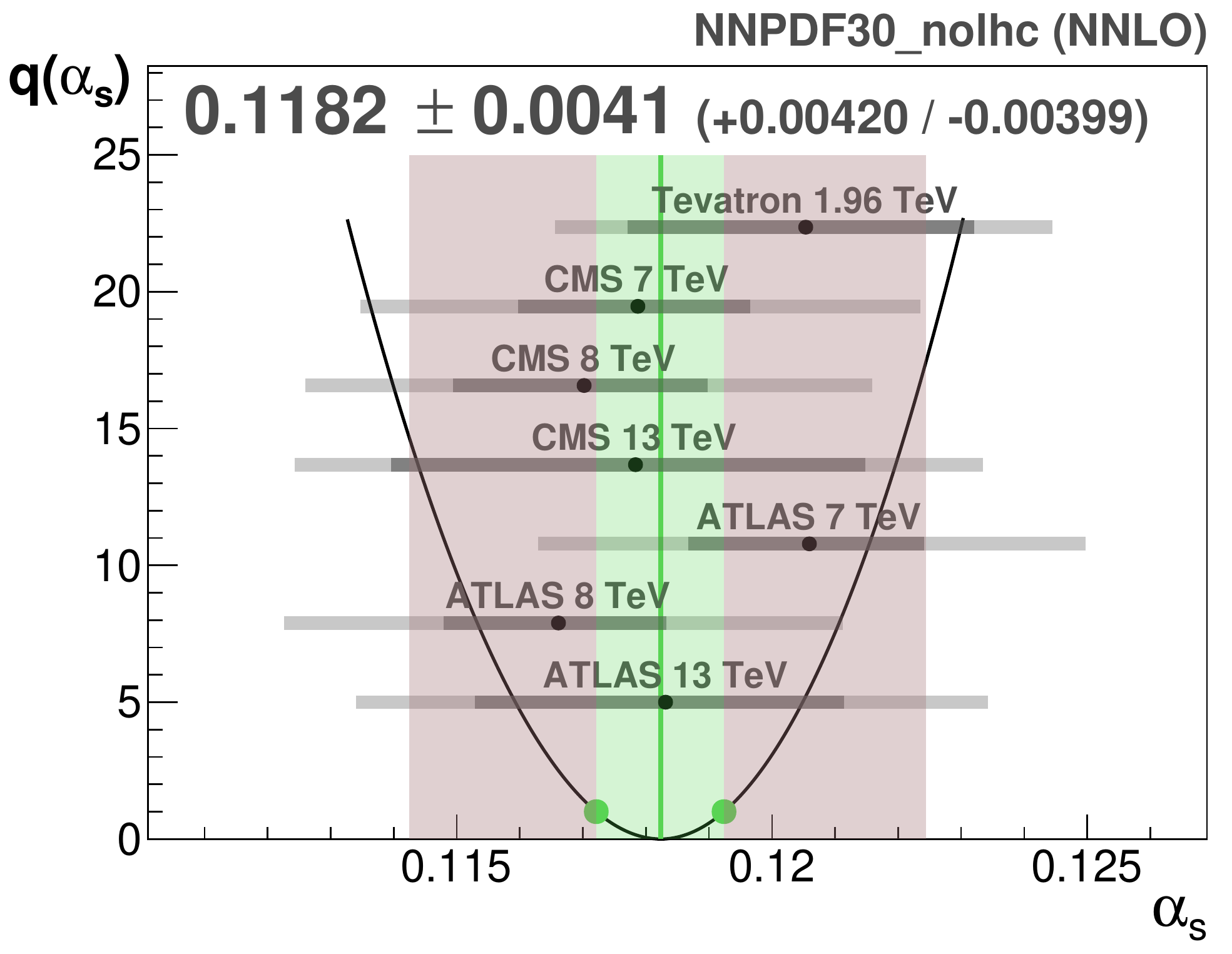%
    }
&
\includegraphics[width=\ScanFigureWidth\linewidth]{%
    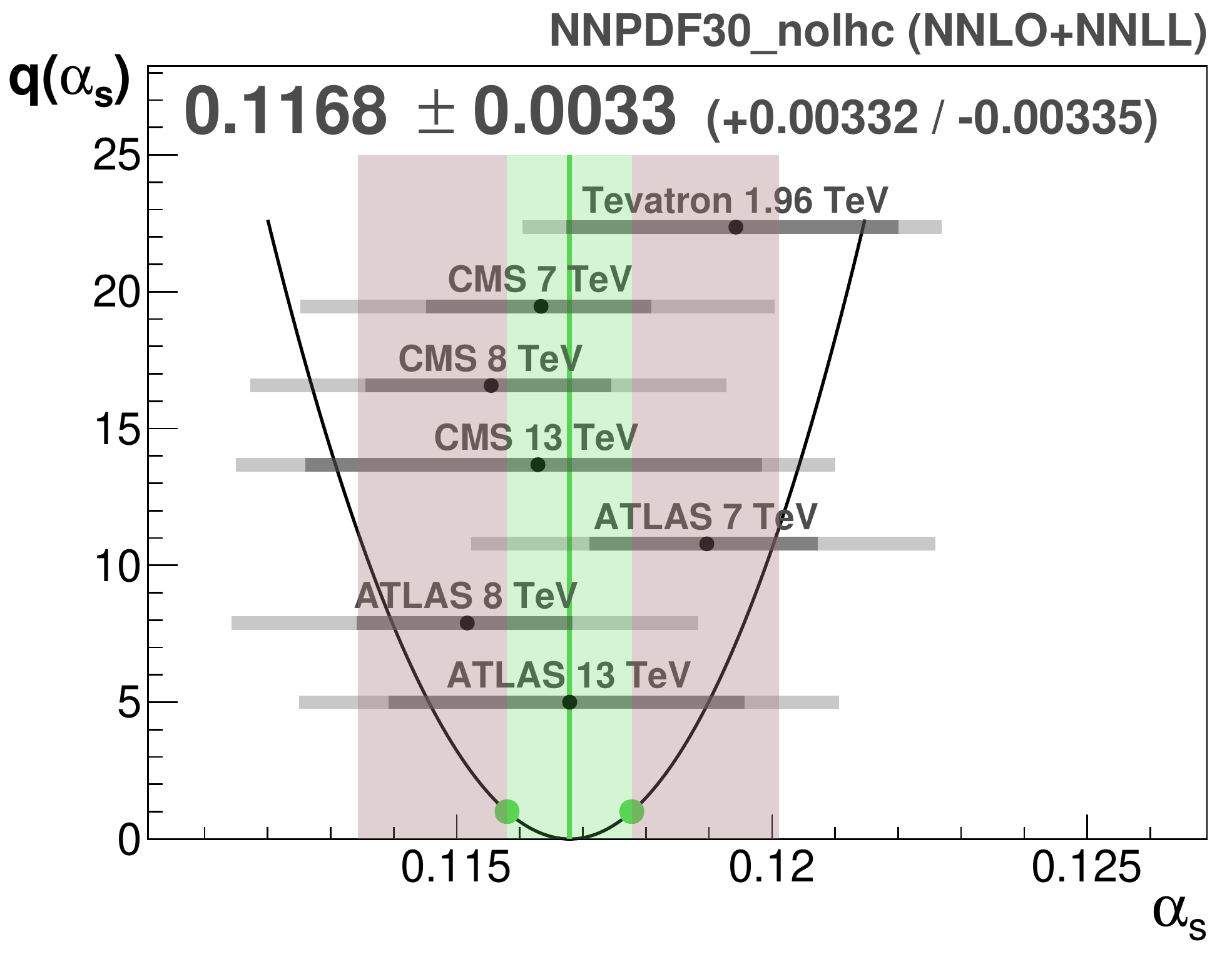%
    }
\\[-6pt]
(c) & (d) \\[6pt]
\end{tabular}
\vspace{-0.3cm}
\caption{
Combination results using the CT14 PDF set (NNLO in (a) and NNLO+NNLL in (b)) and the NNPDF3.0 noLHC PDF set (NNLO in (c) and NNLO+NNLL in (d)).
The individual determinations and their uncertainties are shown in grey, where the darker shade represents the experimental uncertainties which enter into the combination.
The test statistic $q$ as a function of $\as$ is plotted as a black line.
The green line and band represent the central value of the combination and the $1\sigma$ confidence interval respectively. 
The red band depicts the total combination uncertainty with scale, PDF
and top-mass uncertainties included.
}
\label{fig:ScanResults}
\end{figure}
\begin{figure}[htb]
\centering
\includegraphics[width=0.6\linewidth]{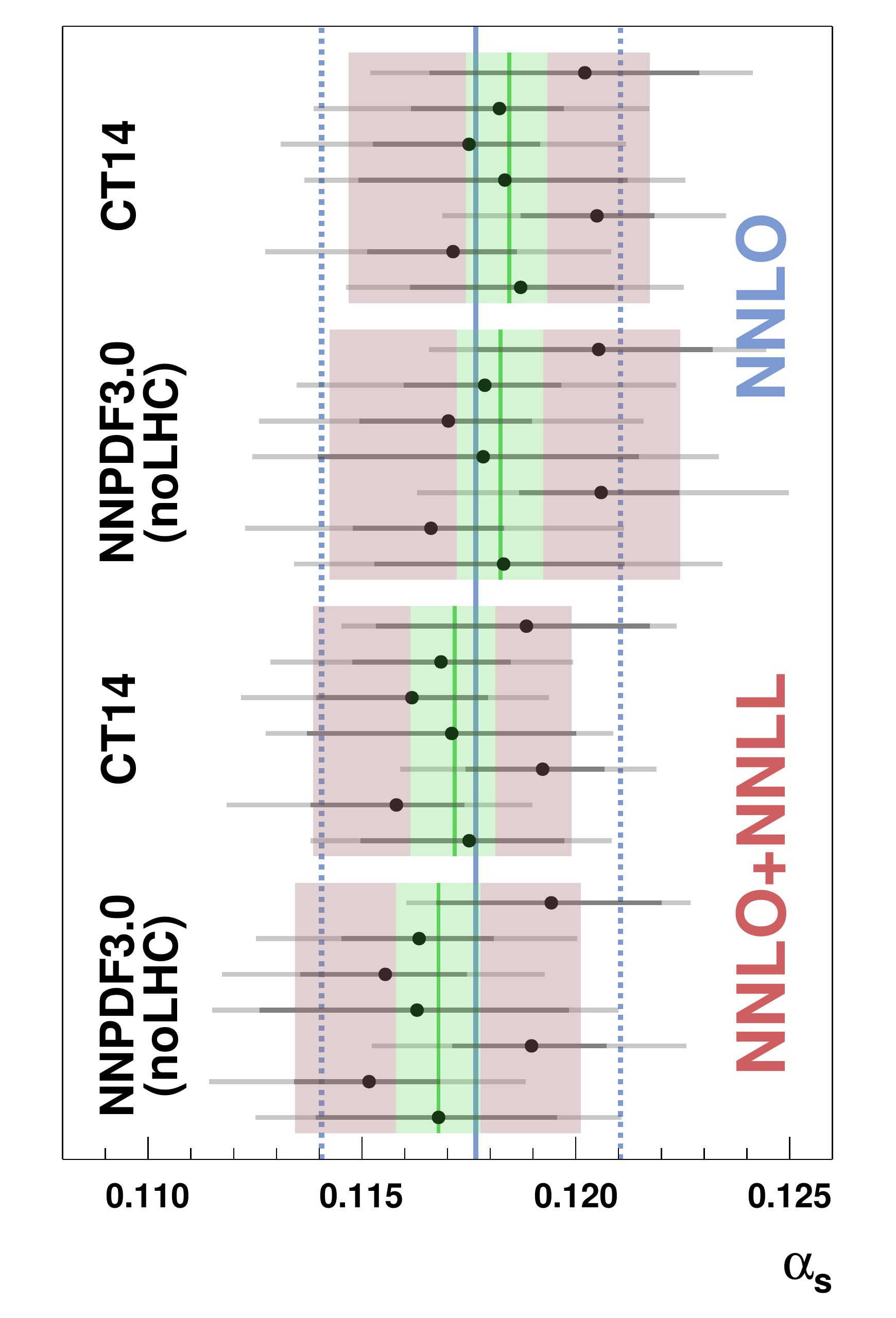}
\vspace{-0.3cm}
\caption{
Combination results for all PDF sets taken into consideration, at NNLO
and NNLO+NNLL. The solid blue line is the unweighted average of the
individual combination results, and the dashed blue lines represent
the 68\% confidence interval.
The red and green bands are as in Fig.~\ref{fig:ScanResults}.
}
\label{fig:unweightedaverage}
\end{figure}
%

\section{Conclusions}
We have used seven measurements of the top-antitop quark production cross section
at the LHC and the Tevatron in order to determine the strong coupling constant $\asmz$,
using the CT14 PDF set and the NNPDF30\_nolhc PDF set at NNLO and NNLO+NNLL.
Overall, our determination of $\as$ yields a value that is compatible
with the world average value and uncertainties that are somewhat larger
than the best individual determinations, though comparable to that
from the electroweak precision data~\cite{Baak:2014ora}.
The largest uncertainties are associated with unknown higher-order
contributions and PDF uncertainties.

\section*{Acknowledgements}

TK is supported by grant Nr. 200020\_162665 of the Swiss National Science Foundation.
GPS is supported in part by ERC Advanced Grant Higgs@LHC (No.\ 321133)
and also wishes to thank the Munich Institute for Astro- and Particle
Physics for hospitality while this work was being completed.
We are grateful to Andreas Jung regarding
discussions of the Tevatron beam-energy uncertainty.

\appendix

\section{Including strongly correlated uncertainty sources in the combination}
\label{sec:appendix}

Our approach of excluding strongly correlated uncertainties from the combination is generally recommended when using the covariance matrix to fit strongly correlated data~\cite{DAgostini:1993arp}.
To illustrate the effect of strong correlations, the combination is here performed again with the PDF, scale and $\mt$ uncertainties included in the combination one by one.
The PDF and scale uncertainties are considered fully correlated between measurements made with the LHC, and partially correlated between measurements made with the LHC and Tevatron.
In the case of the PDF uncertainties the degree of correlation between LHC and Tevatron measurements was determined using the procedure described in Ref.~\cite{LHAPDF}.
The $\mt$ uncertainties are considered fully correlated for all measurements.

Tables~\ref{tab:theoryUncertainties_CT14_NNLO} and \ref{tab:theoryUncertainties_CT14_NNLO_NNLL} show the results for the CT14 PDF set, using NNLO and NNLO+NNLL respectively, and Tables~\ref{tab:theoryUncertainties_NNPDF30_NNLO} and \ref{tab:theoryUncertainties_NNPDF30_NNLO_NNLL} for the NNPDF30\_nolhc PDF set.
As expected, the total uncertainty decreases as more sources are included in the combination.
As the sensitivity to $\as$ is stronger for a larger cross section, determinations that deviate up can have a smaller uncertainty, and therefore obtain a larger weight in the combination.
This is the case for the determination from the ATLAS measurement at 7$\TeV$ when using the CT14 PDF set.
A larger weight may also be obtained for determinations that are more independent with respect to the others. This is primarily the case for the Tevatron determination, for both PDF sets.
These effects are enhanced if the overall correlation is increased by including strongly correlated uncertainty sources, which explains why the combination yields increasing values of $\as$ as more sources are included.
The results found this way are larger than both the straight average and the median of the individual determinations, though the difference is well within one standard deviation. 
Taking them as our final results would imply a high degree of trust in the assumed correlations.
Due to the inherent difficulty of determining correlations, notably as concerns the scale variations, and the importance of the subtle interplay between an individual determination's $\alpha_s$ result and its error, the conservative approach is to exclude the strongly correlated sources from the combination.

\begin{table}[ht] 
\renewcommand{\arraystretch}{1.4}
\setlength\tabcolsep{5pt}
\centering
{\scriptsize
    \hspace*{-0.75cm}%
    \begin{minipage}[t]{0.5\hsize}\centering
    \begin{tabular}{l l l l}
    \toprule
    \lcell{2.4cm}{Uncertainties included in combination} & 
    \lcell{0.7cm}{Center} & 
    \lcell{1.5cm}{Combination uncertainty} & 
    \lcell{1.5cm}{Total uncertainty} \\ 
    \midrule
-                                         &  $0.1184  $  &  ${}_{-0.0010}^{+0.0009}$  &  ${}_{-0.0038}^{+0.0033}$ \\
PDF                                       &  $0.1191  $  &  ${}_{-0.0018}^{+0.0018}$  &  ${}_{-0.0034}^{+0.0033}$ \\
PDF and $\mt$                             &  $0.1194  $  &  ${}_{-0.0020}^{+0.0020}$  &  ${}_{-0.0032}^{+0.0032}$ \\
PDF, $\mt$ and scale                      &  $0.1207  $  &  ${}_{-0.0029}^{+0.0030}$  &  ${}_{-0.0029}^{+0.0030}$ \\
    \bottomrule
    \end{tabular} 
    \caption{\small
        Combination results including also uncertainties from the PDF, the scale
        and the top mass in the combination.
        The first row corresponds to our
        approach of excluding correlated uncertainties from the combination.
        The results listed here are obtained using NNLO cross sections with
        the CT14nnlo series of PDFs.
        }
    \label{tab:theoryUncertainties_CT14_NNLO}
    \end{minipage}%
    \hspace*{1cm}%
    %
    \begin{minipage}[t]{0.5\hsize}\centering
    \begin{tabular}{l l l l }
    \toprule
    \lcell{2.4cm}{Uncertainties included in combination} & 
    \lcell{0.7cm}{Center} & 
    \lcell{1.5cm}{Combination uncertainty} & 
    \lcell{1.5cm}{Total uncertainty} \\ 
    \midrule
-                                         &  $0.1172  $  &  ${}_{-0.0010}^{+0.0010}$  &  ${}_{-0.0033}^{+0.0027}$ \\
PDF                                       &  $0.1180  $  &  ${}_{-0.0020}^{+0.0019}$  &  ${}_{-0.0029}^{+0.0027}$ \\
PDF and $\mt$                             &  $0.1183  $  &  ${}_{-0.0022}^{+0.0022}$  &  ${}_{-0.0027}^{+0.0027}$ \\
PDF, $\mt$ and scale                      &  $0.1188  $  &  ${}_{-0.0025}^{+0.0025}$  &  ${}_{-0.0025}^{+0.0025}$ \\
    \bottomrule
    \end{tabular} 
    \caption{\small
        As in Table~\ref{tab:theoryUncertainties_CT14_NNLO}, but now using NNLO+NNLL
        cross sections with the CT14nnlo series of PDFs.
        }
    \label{tab:theoryUncertainties_CT14_NNLO_NNLL}
    \end{minipage}
    %
    \\[16pt]
    \hspace*{-0.75cm}%
    \begin{minipage}[t]{0.5\hsize}\centering
    \begin{tabular}{l l l l }
    \toprule
    \lcell{2.4cm}{Uncertainties included in combination} & 
    \lcell{0.7cm}{Center} & 
    \lcell{1.5cm}{Combination uncertainty} & 
    \lcell{1.5cm}{Total uncertainty} \\ 
    \midrule
-                                         &  $0.1182  $  &  ${}_{-0.0010}^{+0.0010}$  &  ${}_{-0.0040}^{+0.0042}$ \\
PDF                                       &  $0.1188  $  &  ${}_{-0.0022}^{+0.0023}$  &  ${}_{-0.0037}^{+0.0040}$ \\
PDF and $\mt$                             &  $0.1190  $  &  ${}_{-0.0024}^{+0.0025}$  &  ${}_{-0.0036}^{+0.0040}$ \\
PDF, $\mt$ and scale                      &  $0.1200  $  &  ${}_{-0.0036}^{+0.0035}$  &  ${}_{-0.0036}^{+0.0035}$ \\
    \bottomrule
    \end{tabular} 
    \caption{\small
        As in Table~\ref{tab:theoryUncertainties_CT14_NNLO}, but now using NNLO
        cross sections with the NNPDF30\_nolhc series of PDFs.
        }
    \label{tab:theoryUncertainties_NNPDF30_NNLO}
    \end{minipage}%
    \hspace*{1cm}%
    %
    \begin{minipage}[t]{0.5\hsize}\centering
    \begin{tabular}{l l l l }
    \toprule
    \lcell{2.4cm}{Uncertainties included in combination} & 
    \lcell{0.7cm}{Center} & 
    \lcell{1.5cm}{Combination uncertainty} & 
    \lcell{1.5cm}{Total uncertainty} \\ 
    \midrule
-                                         &  $0.1168  $  &  ${}_{-0.0010}^{+0.0010}$  &  ${}_{-0.0034}^{+0.0033}$ \\
PDF                                       &  $0.1175  $  &  ${}_{-0.0023}^{+0.0023}$  &  ${}_{-0.0031}^{+0.0031}$ \\
PDF and $\mt$                             &  $0.1178  $  &  ${}_{-0.0025}^{+0.0025}$  &  ${}_{-0.0030}^{+0.0030}$ \\
PDF, $\mt$ and scale                      &  $0.1182  $  &  ${}_{-0.0028}^{+0.0028}$  &  ${}_{-0.0028}^{+0.0028}$ \\
    \bottomrule
    \end{tabular} 
    \caption{\small
        As in Table~\ref{tab:theoryUncertainties_CT14_NNLO}, but now using NNLO+NNLL
        cross sections with the NNPDF30\_nolhc series of PDFs.
        }
    \label{tab:theoryUncertainties_NNPDF30_NNLO_NNLL}
    \end{minipage}%
}%
\end{table}

\section{Overview of asymmetric uncertainties used in the combination}
\label{sec:appendix2}

Tables~\ref{tab:uncertaintyCoefficients_CT14_NNLO} and \ref{tab:uncertaintyCoefficients_CT14_NNLO_NNLL} show the numerical values for the uncertainty coefficients used in the combination procedure for the CT14 PDF set, using NNLO and NNLO+NNLL cross sections respectively.
Only experimental uncertainties are listed.
%
Theoretical uncertainties, which are taken into account after the combination procedure, can be found in Tables~\ref{tab:determination_NNLO_CT14}-\ref{tab:determination_NNLO_NNLL_NNPDF30nolhc}.
The correlations for the correlated uncertainties (with a $\delta$ symbol) are described in Section~\ref{sec:correlationcoefficients}.

\begin{table}[ht] 
\renewcommand{\arraystretch}{2.0}
\renewcommand{\ErrTableWidth}{1.0cm}
\setlength\tabcolsep{5pt}
\centering
{\footnotesize
    \hspace*{-0.75cm}%
    \begin{minipage}[t]{0.5\hsize}\centering
        \begin{tabular}{l c c c c c }
        \toprule
Exp.                        & $\delta^{\text{Syst.}}$     & $\delta^{\text{Lumi.}}$     & $\delta^{\text{E}_{beam}}$  & $\Delta^{\text{Stat.}}$     & $\Delta^{\text{Lumi.}}$     \\
\midrule
\cell{\ErrTableWidth}{ATLAS (13\,TeV)}      & ${}_{-0.0021}^{+0.0017}$    & ${}_{-0.0002}^{+0.0002}$    & ${}_{-0.0001}^{+0.0001}$    & ${}_{-0.0006}^{+0.0006}$    & ${}_{-0.0014}^{+0.0012}$    \\
\cell{\ErrTableWidth}{ATLAS (8\,TeV)}       & ${}_{-0.0014}^{+0.0011}$    & ${}_{-0.0004}^{+0.0003}$    & ${}_{-0.0002}^{+0.0001}$    & ${}_{-0.0004}^{+0.0003}$    & ${}_{-0.0013}^{+0.0009}$    \\
\cell{\ErrTableWidth}{ATLAS (7\,TeV)}       & ${}_{-0.0012}^{+0.0009}$    & ${}_{-0.0003}^{+0.0002}$    & ${}_{-0.0001}^{+0.0001}$    & ${}_{-0.0009}^{+0.0007}$    & ${}_{-0.0010}^{+0.0007}$    \\
\cell{\ErrTableWidth}{CMS (13\,TeV)}        & ${}_{-0.0030}^{+0.0025}$    & ${}_{-0.0002}^{+0.0002}$    & ${}_{-0.0001}^{+0.0002}$    & ${}_{-0.0007}^{+0.0006}$    & ${}_{-0.0015}^{+0.0013}$    \\
\cell{\ErrTableWidth}{CMS (8\,TeV)}         & ${}_{-0.0015}^{+0.0011}$    & ${}_{-0.0004}^{+0.0003}$    & ${}_{-0.0001}^{+0.0001}$    & ${}_{-0.0004}^{+0.0003}$    & ${}_{-0.0016}^{+0.0012}$    \\
\cell{\ErrTableWidth}{CMS (7\,TeV)}         & ${}_{-0.0014}^{+0.0010}$    & ${}_{-0.0003}^{+0.0002}$    & ${}_{-0.0002}^{+0.0001}$    & ${}_{-0.0007}^{+0.0005}$    & ${}_{-0.0013}^{+0.0009}$    \\
\cell{\ErrTableWidth}{Tevatron (1.96\,TeV)} & ${}_{-0.0026}^{+0.0019}$    & -                           & -                           & ${}_{-0.0018}^{+0.0013}$    & ${}_{-0.0019}^{+0.0014}$    \\
        \bottomrule
        \end{tabular} 
        \caption{\small
            Overview of the uncertainty coefficients for the CT14 (NNLO) PDF set.
            The coefficients with a $\delta$ correspond to the coefficients of the correlated
            uncertainty sources, and those with a $\Delta$ to the uncorrelated uncertainty sources.
            }
        \label{tab:uncertaintyCoefficients_CT14_NNLO}
        \end{minipage}%
    \hspace*{1cm}%
    %
    \begin{minipage}[t]{0.5\hsize}\centering
        \begin{tabular}{l c c c c c }
        \toprule
Exp.                        & $\delta^{\text{Syst.}}$     & $\delta^{\text{Lumi.}}$     & $\delta^{\text{E}_{beam}}$  & $\Delta^{\text{Stat.}}$     & $\Delta^{\text{Lumi.}}$     \\
\midrule
\cell{\ErrTableWidth}{ATLAS (13\,TeV)}      & ${}_{-0.0020}^{+0.0018}$    & ${}_{-0.0002}^{+0.0002}$    & ${}_{-0.0001}^{+0.0001}$    & ${}_{-0.0006}^{+0.0005}$    & ${}_{-0.0014}^{+0.0012}$    \\
\cell{\ErrTableWidth}{ATLAS (8\,TeV)}       & ${}_{-0.0014}^{+0.0011}$    & ${}_{-0.0004}^{+0.0003}$    & ${}_{-0.0002}^{+0.0001}$    & ${}_{-0.0004}^{+0.0004}$    & ${}_{-0.0013}^{+0.0010}$    \\
\cell{\ErrTableWidth}{ATLAS (7\,TeV)}       & ${}_{-0.0012}^{+0.0010}$    & ${}_{-0.0003}^{+0.0002}$    & ${}_{-0.0001}^{+0.0001}$    & ${}_{-0.0009}^{+0.0007}$    & ${}_{-0.0010}^{+0.0008}$    \\
\cell{\ErrTableWidth}{CMS (13\,TeV)}        & ${}_{-0.0029}^{+0.0025}$    & ${}_{-0.0002}^{+0.0002}$    & ${}_{-0.0002}^{+0.0001}$    & ${}_{-0.0007}^{+0.0006}$    & ${}_{-0.0015}^{+0.0013}$    \\
\cell{\ErrTableWidth}{CMS (8\,TeV)}         & ${}_{-0.0015}^{+0.0012}$    & ${}_{-0.0004}^{+0.0003}$    & ${}_{-0.0002}^{+0.0001}$    & ${}_{-0.0004}^{+0.0003}$    & ${}_{-0.0016}^{+0.0012}$    \\
\cell{\ErrTableWidth}{CMS (7\,TeV)}         & ${}_{-0.0015}^{+0.0011}$    & ${}_{-0.0003}^{+0.0002}$    & ${}_{-0.0002}^{+0.0001}$    & ${}_{-0.0007}^{+0.0006}$    & ${}_{-0.0013}^{+0.0010}$    \\
\cell{\ErrTableWidth}{Tevatron (1.96\,TeV)} & ${}_{-0.0025}^{+0.0021}$    & -                           & -                           & ${}_{-0.0017}^{+0.0014}$    & ${}_{-0.0018}^{+0.0015}$    \\
        \bottomrule
        \end{tabular} 
        \caption{\small
            As in Table~\ref{tab:uncertaintyCoefficients_CT14_NNLO}, but now using
            NNLO+NNLL cross sections with the CT14 PDF set.
            }
        \label{tab:uncertaintyCoefficients_CT14_NNLO_NNLL}
        \end{minipage}
}%
\end{table}

\end{document}